\documentclass[pdflatex,sn-mathphys-num]{sn-jnl}
\usepackage{graphicx}%
\usepackage{multirow}%
\usepackage{amsmath,amssymb,amsfonts}%
\usepackage{amsthm}%
\usepackage{mathrsfs}%
\usepackage[title]{appendix}%
\usepackage[dvipsnames]{xcolor}
\usepackage{textcomp}%
\usepackage{manyfoot}%
\usepackage{booktabs}
\usepackage{algpseudocode}%
\usepackage{listings}%
\usepackage{mathrsfs}
\usepackage[table]{xcolor}
\usepackage{colortbl}
\usepackage{array}
\usepackage{caption}
\usepackage{soul}
\usepackage{booktabs}
\usepackage{tabularx}
\usepackage[table]{xcolor}
\usepackage[makeroom]{cancel}

\usepackage[normalem]{ulem}
\newcommand{\stkout}[1]{\ifmmode\text{\sout{\ensuremath{#1}}}\else\sout{#1}\fi}

\newsavebox\foobox
\newlength{\foodim}
\newcommand{\slantbox}[2][0]{\mbox{%
        \sbox{\foobox}{#2}%
        \foodim=#1\wd\foobox
        \hskip \wd\foobox
        \hskip -0.5\foodim
        \pdfsave
        \pdfsetmatrix{1 0 #1 1}%
        \llap{\usebox{\foobox}}%
        \pdfrestore
        \hskip 0.5\foodim
}}
\def\Laplace{\slantbox[-.45]{$\mathscr{L}$}}

\theoremstyle{thmstyleone}%
\theoremstyle{thmstyletwo}%

\theoremstyle{thmstylethree}%

\begin{document}

\title[Article Title]{Hyperstatistics}

\author[1]
{\fnm{Lucas} \sur{Squillante}}

\author[1]
{\fnm{Samuel M.} \sur{Soares}}

\author[2,3,4,5]{\fnm{Constantino} \sur{Tsallis}}

\author*[1]{\fnm{Mariano} \sur{de Souza}}\email{mariano.souza@unesp.br}

\affil[1]
{\orgdiv{IGCE -- Physics Department}, \orgname{S\~ao Paulo State University (Unesp)}, \orgaddress{\city{Rio Claro}, \state{SP}, \country{Brazil}}}

\affil[2]{\orgname{Centro Brasileiro de Pesquisas Físicas, Rua Dr. Xavier Sigaud 150, Rio de Janeiro 22290-180, Brazil}}

\affil[3]{\orgname{Santa Fe Institute}, \orgaddress{\street{1399 Hyde Park Road}, \city{Santa Fe}, \state{NM}, \postcode{87501}, \country{USA}}}

\affil[4]{\orgname{Complexity Science Hub Vienna}, \orgaddress{\street{Metternichgasse 8}, \city{Vienna}, \postcode{1030}, \country{Austria}}}

\affil[5]{\orgname{Dipartimento di Fisica e Astronomia Ettore Majorana, University of Catania, Via S. Sofia 64, 95123 Catania, Italy}}

\abstract{
We propose a general approach, named by us hyperstatistics, to treat complex systems, in which Boltzmann-Gibbs statistics breaks down at the domain level of the system. Hyperstatistics preserves the concavity of the nonadditive $q$-entropy.
We obtain analytical closed-form expressions for the here proposed ($q, n$)-generalized Boltzmann factor $B_q^{n}$ considering uniform, $\gamma$, Log-normal, F, and the $q$-$\gamma$ probability distribution functions.  Remarkably, for all investigated distribution functions, $B_q^{n}$ reduces to a $q$-exponential-type function.
To demonstrate the applicability of hyperstatistics, we use a table top experiment of the discharge of a capacitor considering $\gamma$-distributed relaxation times, the pressure decay over time associated with the pumping of $^4$He lines of a closed cycle cryostat, midrapidity data for \emph{p}-Pb collisions at the LHC,  as well as data set for acceleration distribution in turbulent systems. Furthermore, we deduce the power-law-like \emph{universal} dielectric response using a $q$-generalized $\gamma$-distribution function proposed in this work.  Hyperstatistics is applicable to systems with inherent non-Boltzmann-Gibbsian statistics at the domain level. We demonstrate that at the domain level hyperstatistics and superstatistics are equivalent, while at the macroscopic level hyperstatistics recovers superstatistics for $q \to 1$ because the Boltzmann-Gibbs character of the domains is restored, as well as Boltzmann-Gibbs statistics for $n \to \infty$.}



\maketitle

\section{Introduction}\label{sec1}
The Boltzmann-Gibbs (BG) statistical framework is widely employed in systems at equilibrium and where short-range forces are dominant. As pointed out by Cohen \cite{Cohen2005}, ``... \emph{The first classical non-Boltzmann statistics of physical systems was proposed by Tsallis} ...'' aiming to treat a variety of complex systems,  cf.\,Refs.\,\cite{tsallis1988,Tsallisbook} and references cited therein. When dealing with systems that lie in nonequilibrium, the approach of superstatistics has been employed, which considers fluctuating physical quantities, such as temperature, that are distributed throughout the system \cite{beck2003}. Even though $q$-statistics and superstatistics have been employed successfully in the description of a variety of systems, cf.\,Refs.\,\cite{Tsallisbook,beck2004,tsallisbeck2025,tsallisarxiv2026}, a general formalism capable of addressing complex systems in terms of probability distribution functions, hereafter PDFs, with high accuracy and mathematical simplicity is still lacking. In the present work, we propose such a new approach, which encapsulates situations whose BG equilibrium in domains of the system is no longer valid, giving rise to what we named as hyperstatistics. The latter was chosen because it is reminiscent of superstatistics and also because of the unprecedented  high accuracy and versatility in the description of experimental data with inherent non-Boltzmann-Gibbsian statistics. Also, as discussed in Section\,\ref{macroscopicBH}, the structure of hyperstatistics at the macroscopic level naturally gives rise to \emph{hyper}geometric functions. Our proposal is not to be confused with a generalization of superstatistics. As a matter of fact, it is distinct from  superstatistics in the sense that we consider a distribution of Boltzmann factors within domains of the system, cf.\,detailed discussions below. Of course, the superstatistics approach also provides a very good description of a variety of systems. Recently, superstatistics was successfully employed to describe chemical reactions as a function of the density of reactants \cite{tsallisbeck2025}. Also, it is remarkable that superstatistics, proposed over 20 years ago, also captures and describes properly the behaviour of turbulent systems \cite{beck2004}. As a matter of fact, superstatistics served as the ground of the present work, since the authors in Ref.\,\cite{beck2003} mention  ``... $f(\beta)$ \emph{cannot be any function but must be a normalized probability density. It may, in fact, be a physically relevant density from statistics, say Gaussian, uniform, chi-squared, lognormal, but also other, as yet unidentified, distributions could be considered if the underlying dynamics is sufficiently complex.}'' As it will be discussed in Section\,\ref{proposal}, the use of the $\gamma$-distribution, cf.\,Eq.\,6 of Ref.\,\cite{wilk2000}, in the obtaining of the generalized Boltzmann factor is in line with Beck and Cohen's statement. It is to be noted that Beck and Cohen deduced the generalized Boltzmann factor in terms of a series expansion, while hyperstatistics is grounded on analytical, exact, closed-form mathematical expressions for the here-proposed ($q, n$)-generalized Boltzmann factor, hereafter $B_q^n$. The use of the term hyperstatistics is also not intended to cover a generalization of statistical mechanics, cf.\,proposed in Ref.\,\cite{tsallis1988}. Instead, as it will be discussed below, hyperstatistics is fully aligned with the $q$-entropic functional at the domain level \cite{tsallis1988} (Supplementary Section 1).  To introduce our proposal, we start by discussing its genesis in terms of the Gamma function $\Gamma(n)$, which is addressed in the following. The classical $\Gamma(n)$ function appears frequently in Physics as the mathematical ground underlying the normalization and computation of exponential-type distribution functions. It turns out that when the latter  are replaced by power-law-type distributions, as required by systems with inherent long-range interaction, disorder, or strong fluctuations, the canonical $\Gamma(n)$ function is no longer appropriate to be used. This motivates the proposal of a ``$q$-deformed'' $\Gamma(n)$ function, where $q$ refers to the entropic index \cite{tsallis1988}. It is to be noted that proposals of generalization of the $\Gamma(n)$ function have been reported in the literature in the form of $q$-analogues associated with $q$-series and hypergeometric functions \cite{andrews}, as well as degenerated $\Gamma(n)$ function \citep{kim}, among others \cite{nadaraja, queiros}. As discussed into more details below, the here proposed $q$-generalized $\Gamma(n)$ function, i.e., $\Gamma(n,q)$, is distinct, since we have defined it as the Mellin transform of the $q$-exponential function. We propose thus a natural generalization of  $\Gamma(n)$ in terms of the $q$-exponential function, hereafter $q$-exp, with the preservation of the mathematical structural role played by $\Gamma(n)$. We start by recalling the $\Gamma(n)$ function, defined as follows \cite{andrews, ricieri, spiegel1963}:
\begin{equation}
\Gamma(n) = \int_0^\infty x^{n-1} e^{-x} \, dx,\label{Gammadefinition}
\end{equation}
which is convergent for $n > 0$. As highlighted by Andrews and Colleagues in Ref.\,\cite{andrews}: ``... \textit{Euler discovered the gamma function, $\Gamma(x)$ function,  when he extended the domain of the factorial function. Thus $\Gamma(x)$ is a meromorphic function equal to $(x - 1)!$ when $x$ is a positive integer.} ...". In the frame of BG statistical mechanics and superstatistics \cite{beck2003}, the exponential function is pivotal, since the  Boltzmann factor is key in computing probabilities. However, in the statistical mechanics framework proposed in Ref.\,\cite{tsallis1988}, the ordinary exponential is replaced by the $q$-exp function, namely:
\begin{equation}
\exp_q(-x) = \big[1 + (q-1)x\big]^{\frac{1}{1-q}}.
\label{qexp}
\end{equation}
Equation \ref{qexp} constitutes a generalization of the exponential function, being its variants  employed successfully in various contexts, see, e.g., Ref.\,\cite{Tsallis2009} and references cited therein. 
We highlight that the origin of the present work lies in the following question: how can we replace the canonical exponential by a $q$-exp function in Boltzmann factor obeying the constraints of statistical mechanics? The answer to this question is encoded in the genesis of hyperstatistics.  
Essentially, in systems at thermal equilibrium, the standard Boltzmann factor provides an appropriate probabilistic description \cite{kittelthermal}. However, in complex systems, where, for instance, a distribution of temperatures within the system takes place, the Boltzmann factor fails to accurately capture the underlying statistics. In such cases, a modified framework is required, and this is precisely where hyperstatistics plays a key role, since it provides a precise statistical description in terms of $B_q^n$.
Before presenting the core of the present work, i.e., the proposal of hyperstatistics, in the following we discuss the $q$-generalized $\Gamma(n)$ function, i.e., $\Gamma(n,q)$, as the Mellin transform \cite{mellin, gradsteyn} of the $q$-exp function. 

\section{Mellin transform of the $q$-exp: $\Gamma(n,q)$, recurrence relation, and convergence conditions}\label{Gammona}
\begin{figure}[!t]
\centering
\includegraphics[width=\textwidth]{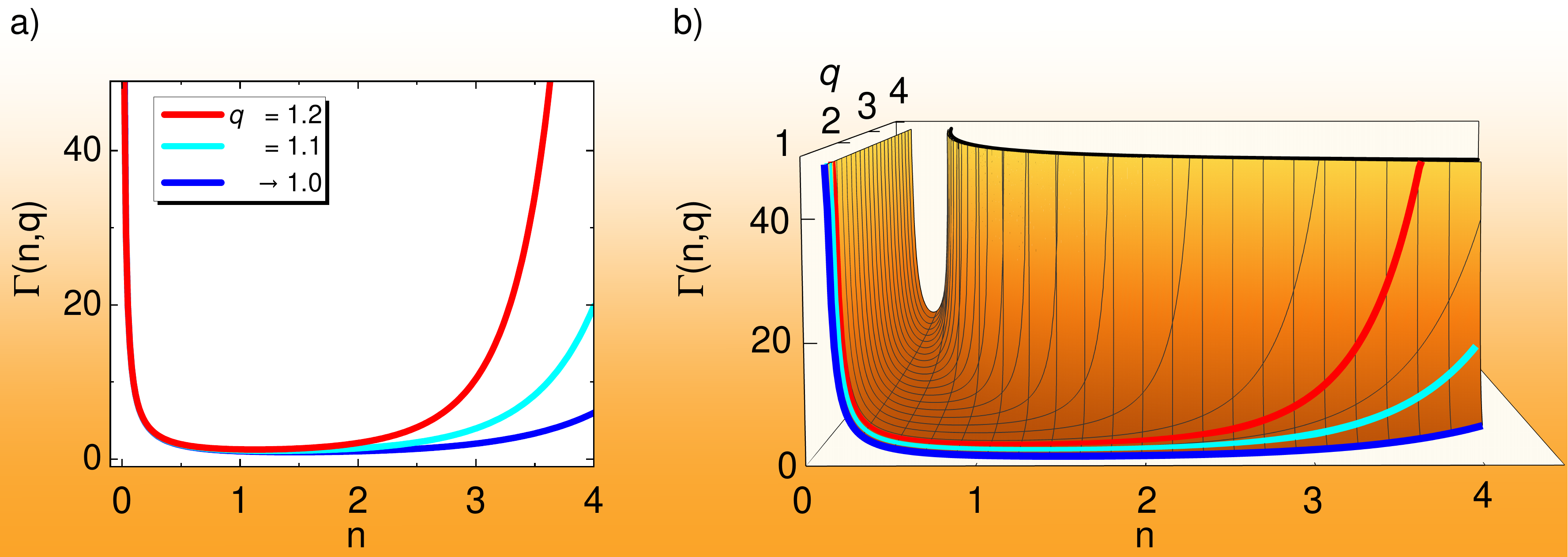}
\caption{\footnotesize \textbf{The $q$-generalized Gamma function $\Gamma(n,q)$}. a)  $\Gamma(n,q)$ $versus$ $n$ for $q$ = 1.2, 1.1, and $q \rightarrow 1$ (canonical $\Gamma(n)$ function, cf.\,Eq.\,\ref{Gammadefinition}). b) $\Gamma(n,q)$ \emph{versus} $n$ \emph{versus} $q$. Note that $\Gamma(n,q)$ is convergent only for $q < (1 + 1/n)$, cf.\,Eq.\,\ref{inequality}, being the upper bound illustrated by the black solid line.}
\label{Fig-1}
\end{figure}

The Mellin transform $F(u)$ of an arbitrary function $\Phi(x)$ reads \cite{mellin}:
\begin{equation}
F(u) = \int_{0}^{\infty} \Phi(x)x^{\,n-1}dx.
\label{Mellintransform}
\end{equation}
Upon considering $\Phi(x) = \exp_q(-x)$, we straightforwardly achieve:
\begin{equation}
F(u) = \int_0^\infty \exp_q(-x) \, x^{n-1}  dx 
= \int_0^\infty \big[1+(q-1)x\big]^{\frac{1}{1-q}} \, x^{n-1}  dx = \Gamma(n,q),
\label{GammaMellin}
\end{equation}
with $n>0$ and $q>1$. The latter is particularly relevant in the frame of hyperstatistics, cf.\,discussions in Section 2 of the Supplementary Material. Although a proposal of a $q$-generalized $\Gamma(n)$ was already reported in Refs.\,\cite{gamma1,gamma2} and the Mellin transform of the $q$-exp function was discussed in Ref.\,\cite{mellinqexp}, to the best of our knowledge the explicit connection between these two quantities, namely the identification of the Mellin transform of the $q$-exp function with $\Gamma(n,q)$, was not been established so far.  This is an important aspect in order to link the Mellin transform of the $q$-exp function with $\Gamma(n,q)$. Now, we make the change of variables $v = (q-1)x$, so that we have:
\begin{equation}
\Gamma(n,q) = \frac{1}{(q-1)^n} \int_0^\infty v^{n-1} (1+v)^{-\frac{1}{q-1}} dv.
\end{equation}
Making use of the mathematical relation between $\Gamma(n)$ and the Beta function \cite{andrews, spiegel1963}:
\begin{equation}
\label{Gamalegal}
\int_0^\infty v^{a-1}(1+v)^{-b} dv
= \frac{\Gamma(a)\Gamma(b-a)}{\Gamma(b)},
\end{equation}
which is valid for $a, b,$ and $(b - a) > 0$, we obtain:
\begin{equation}
\Gamma(n,q) =
\frac{1}{(q-1)^n}
\frac{\Gamma(n)\Gamma\!\left(\frac{1}{q-1}-n\right)}
{\Gamma\!\left(\frac{1}{q-1}\right)},
\label{Gamma nq}
\end{equation}
with:
\begin{equation}
\left(\frac{1}{q-1} - n \right) > 0,
\label{inequality}
\vspace{0.3cm}
\end{equation}
aiming to assure integrability for $x \rightarrow 0$, whereas the typical power-law-type decay, i.e., $\exp_q(-x)\sim x^{-1/(q-1)}$ for $x \rightarrow \infty$, imposes $q < (1 + 1/n)$ as the upper bound, so that $q \in (1, 1+1/n)$, cf.\,Fig.\,\ref{Fig-1}. It is worth emphasizing that the convergence criterion in the definition of $\Gamma(n,q)$, cf.\,Eqs.\,\ref{GammaMellin} and \ref{inequality}, is key in the applications of hyperstatistics to be discussed in Section \ref{applications}. The limit $\exp_q(-x)\sim x^{-1/(q-1)}$ for $x \rightarrow \infty$ shall be particularly relevant to treat dielectric response functions in the low-frequency regime, cf.\,discussions in Section \ref{dielectric}.
Note that the definition of an upper bound for the convergence criterion of $\Gamma(n,q)$ is tricky. This is because upon considering integer values of $n$, for instance, $n = 1 \Rightarrow q < 2$. However, for fractionary values of $n$, for instance, $n = 1/2$  $\Rightarrow q < 3$. The details of the convergence criterion (Eq.\,\ref{inequality}) can be better visualized in Fig.\,\ref{Fig-1}. 
It is remarkable that, in our approach, we obtain an interval of validity for the values of $q$ defined by the bounds imposed by Eq.\,\ref{inequality}. The latter is distinct from the results reported in Ref.\,\cite{ourabah}, where the possible values of $q$ are constrained by an equality.
It turns out that the Mellin transform of the $q$-exp function, i.e., $\Gamma(n,q)$, naturally enables us to connect it with the $Beta$ function \cite{ricieri} in a more general way, namely:
\begin{equation}
\Gamma(n,q)=\int_0^\infty x^{n-1}\exp_q(-x)\,dx = \frac{1}{(q-1)^n}
Beta\!\left(n,\frac{1}{q-1}-n\right).
\label{gammabeta}
\end{equation}
Also, a well-defined property of the classical $\Gamma(n)$ function is the recurrence relation $\Gamma(n+1) = n\,\Gamma(n)$ \cite{ricieri}. For $\Gamma(n,q)$, it is straightforward to obtain:
\begin{equation}
\Gamma(n+1,q)
=
\frac{n}{[1-(q-1)(n+1)]}\,
\Gamma(n,q),
\end{equation}
which is valid in the convergence domain $q \in (1, 1+1/n)$ with $n > 0$. It is evident that for $q \to 1$, the classical recurrence relation is nicely recovered. Also, it is noteworthy that $\lim_{q\to 1} \Gamma(n,q) = \Gamma(n)$ analogously to the fact that the $q$-exp function reduces to the ordinary exponential function in the same limit. In its quintessence, $\Gamma(n, q)$ provides a smooth deformation of the classical $\Gamma(n)$ function controlled by $q$, cf.\,illustrated in Fig.\,\ref{Fig-1}. Thus, $\Gamma(n, q)$ can be seen as a practical analytical \textit{tool} for the normalization of $q$-exp-type distribution functions. In this context, we propose a $q$-$\gamma$ PDF, in terms of $\Gamma(n,q)$, cf.\,discussions in Sections \ref{proposal} and \ref{applications}, and Table\,\ref{Table-1}. Note that the emergence of $\Gamma(n, q)$ comes from the Mellin transform of the $q$-exp, cf.\,Eq.\,\ref{GammaMellin}. Next, we introduce our proposal of hyperstatistics.

\section{Proposal of hyperstatistics}\label{proposal}
In their seminal work on superstatistics \cite{beck2003}, Beck and Cohen considered a system in nonequilibrium with a fluctuating intensive observable across different domains of the system. They considered the Laplace transform of the PDF as the generalized Boltzmann factor, being argued by them that $q$-statistics is a particular case of their approach in terms of PDFs at the macroscopic level \cite{wilk2000,beckprl,beck2003}. As it shall be discussed below, our proposal of hyperstatistics is grounded on the consideration of a $\gamma$-distribution of Boltzmann factors within a domain of the system, cf.\,illustrated in Fig.\,\ref{Fig-2}, which astonishingly enables us to obtain $B_q^n$ in closed form in terms of $q$-exp for all here investigated PDFs. 
\begin{figure}[t!]
\centering
\includegraphics[width=1.00\textwidth]{Fig-2.pdf}
\caption{\footnotesize \textbf{Non-Boltzmann-Gibbsian statistics within each domain composing the system: hyperstatistics.} a) Schematic representation of domains with distinct inverse temperatures $\beta_i$ for a macroscopic system. Each domain encapsulates a $\gamma$-distribution of Boltzmann factors, e.g., $e^{-\beta_1' E}$, $e^{-\beta_1'' E}$, \ldots, cf.\,indicated by the white arrow in the domain comprising the inverse temperature $\beta_1$. $\langle \beta \rangle$ refers to the mean-value of $\langle\beta_i\rangle$ for the system. b) The system is composed by a distribution of $e_q^{-\langle \beta_i \rangle E}$ factors, cf.\,indicated by the black arrow, i.e., non-Boltzmann-Gibbsian statistics in the domain level. c) $q$-exp-based Boltzmann factors in the domains lead to a unique $e_q^{-\langle \beta_i \rangle E}$, being $\langle \beta_i \rangle$ associated with the measured temperature in a real system \cite{afewremarks}. Note that the expected value of $\exp_q(-\langle \beta_i \rangle E)$, i.e., $\langle \exp_q(-\langle \beta_i \rangle E) \rangle$, describes the system at the macroscopic level, cf.\,discussions in Section\,\ref{macroscopicBH}.
Figure adapted from Ref.\,\cite{cbeckfig}. d) The canonical Boltzmann factor for distinct $\beta_i$ inside a given domain, i.e., $\beta_1'$, $\beta_1''$, $\beta_1'''$ (dotted lines). Upon employing the Laplace transform of the $\gamma$-distribution, the non-Boltzmann-Gibbsian statistics in the level of domains, i.e., $q$-exp function, naturally emerges (solid line). e) Measuring temperature in a real system provides experimental access to $\langle \beta_i \rangle$, which is the average of slightly distinct temperatures in the surrounding of the thermometer. Dotted lines depict distinct $q$-generalized Boltzmann weights \emph{versus} $E$ that result in a single ($q, n$)-generalized Boltzmann factor $B_q^{n}$ (solid line) that incorporates a $q$-exp function in terms of $\langle \beta_i \rangle$.  f) The flowchart indicates the steps to obtain, cf.\,panels a), b), and c), $B_q^{n}$ upon considering a $\gamma$-distribution to compute $\langle \beta_i\rangle$. Details in the main text.}
\label{Fig-2}
\end{figure}
It is worth emphasizing that, although the intensive parameter $\beta_i$ indeed fluctuates, each value of $\beta_i$ is associated with a corresponding Boltzmann weight, as depicted in Fig.\,\ref{Fig-2} a). Note that our consideration of a distribution of Boltzmann factors associated with distinct $\beta_i$s is based on the mathematical structure of the $\gamma$-distribution, which in turn comes from a sum of Poisson distributions, cf.\,discussions in the Supplementary Section 3. Hence, our approach is applicable to systems characterized by a distribution of non-Boltzmann-Gibbsian microstates within domains composing the whole system. As a step further, we then extend hyperstatistics to the macroscopic level and make a direct comparison with superstatistics for the $\gamma$-distribution.

\subsection{Microscopic and mesoscopic (domain) levels}
We start by considering a $\gamma$-distribution, cf.\,Eq.\,6 of Ref.\,\cite{wilk2000}, of the Boltzmann factor, i.e., a \emph{genuine} Laplace transform of the $\gamma$-distribution, as follows (Supplementary Section 4):
\begin{equation}
\Laplace\{\gamma(\beta_i)\} = \frac{1}{\Gamma(n)}\left(\frac{n}{\langle \beta_i \rangle}\right)^{n}
\int_{0}^{\infty} e^{-\beta_i E}\beta_i^{n-1}e^{-\frac{n\beta_i}{\langle \beta_i \rangle}}d\beta_i,
\label{Bfactor}
\end{equation}
where $\beta_i$ is the inverse temperature of the i\emph{th}-domain and $E$ the Hamiltonian Eigenenergy, cf.\,Fig.\,\ref{Fig-2}, and the inverse temperature
$\beta$ for the entire system is computed as $\langle \beta \rangle = \int_{0}^{\infty} \langle\beta_i\rangle \gamma(\langle\beta_i\rangle)d\langle\beta_i\rangle$ \cite{beck2003}.  Note that $\langle \beta_i \rangle$ refers to the average inverse temperature of a given domain $i$ considering a distribution of temperatures within such a domain, cf.\,Figs.\,\ref{Fig-2} a) and b). Solving the integral in Eq.\,\ref{Bfactor} and making simplifications, we have:
\begin{equation}
\Laplace\{\gamma(\beta_i)\} = \left(\frac{n}{\langle \beta_i \rangle}\right)^{n} \frac{1}{\left(E + \frac{n}{\langle \beta_i \rangle}\right)^n}.
\label{Bfactor1}
\end{equation}
Rearranging Eq.\,\ref{Bfactor1} and making the substitution $n = -\frac{1}{1-q}$, we obtain:
\begin{equation}
\Laplace\{\gamma(\beta_i)\} = \left(1+\frac{\langle \beta_i \rangle E}{n}\right)^{-n} = \left[1+(q-1)\langle \beta_i \rangle E\right]^{\frac{1}{1-q}} = e_q^{-\langle \beta_i \rangle E},
\label{finallaplace}
\end{equation}
i.e.,  the $\gamma$-distribution of $\beta_i$, associated with $e^{-\beta_i E}$, leads naturally to a $e_q^{-\langle \beta_i \rangle E}$, cf.\,Figs.\,\ref{Fig-2} a) and b). It is clear that the latter embodies a power-law in terms of $q$, cf.\,Eq.\,\ref{qexp}.
This deduction is possible because only the $\gamma$-distribution intrinsically embodies a sum of exponential weights associated with the Poisson distribution, cf.\,Eqs.\,(S30) and (S41) in Section 3 of the Supplementary Material.
 Remarkably, the term $\left(1+\frac{\langle \beta_i \rangle E}{n}\right)^{-n}$ in Eq.\,\ref{finallaplace} can be recognized as an $n$-exp function, i.e., exp$_n\left(-\langle \beta_i \rangle E\right)$, which recovers $\exp{(-\langle \beta_i \rangle E)}$ in the limit $n \rightarrow \infty$. A detailed discussion of the consequences of using an $n$-exp function is presented in Section 2 of the Supplementary Material. In the present work, we stick to the representation of the result of the Laplace transform of the $\gamma$-distribution in terms of a $q$-exp function. Note that $\Laplace\{\gamma(\beta_i)\}$ gives rise naturally to the $q$-exp function, which is employed in the computation of $B_q^n(E)$ in terms of both $n$ and $q$. This is a key point associated with the accuracy of hyperstatistics. We propose thus that upon considering a $\gamma$-distribution of $\beta_i$ at the domain level, the usual exponential term can be naturally replaced by a $q$-exp function comprising $\langle \beta_i \rangle$ in its argument, so that the here proposed $B_q^{n}$, applicable at the domain level, reads:
\begin{equation}
\boxed{B_q^{n} (E) = \exp_q(-\langle \beta_i \rangle E)\, = \exp_q\left[-\left(\int_0^{\infty}\beta_i q\textmd{-}\gamma(\beta_i)d(\beta_i)\right)E\right],}
\label{qBfactor}
\end{equation}
\noindent where $\langle \beta_i \rangle = \int_0^{\infty}\beta_i q\textmd{-}\gamma(\beta_i) d(\beta_i) = \frac{\beta_0}{[1+(1-q)(n+1)]}$, and $q\textmd{-}\gamma(\beta_i) = \frac{1}{\Gamma(n,q)}\left(\frac{n}{\beta_0}\right)^n{\beta_i}^{n-1}\exp_q{\left(-\frac{n\beta_i}{\beta_0}\right)}$ (Supplementary Section 5) is the here-proposed $q$-generalized PDF $\gamma$-distribution. Interestingly, for $q \rightarrow 1$ $\Rightarrow$ $\langle \beta_i \rangle \rightarrow \beta_0$, which recovers $\beta_0 = \int_0^{\infty}\beta f(\beta)d\beta$ in the frame of superstatistics \cite{beck2003}. From the experimental point of view, this is directly connected to the measurement of temperature in a real system because one cannot measure the temperature of the system as a whole \cite{reviewmariano}. As a matter of fact, attaching a thermometer to a sample yields the average temperature only over a domain of the specimen, cf.\,Fig.\,1 of Ref.\,\cite{afewremarks}. This is the reason why $B_q^n(E)$ is expressed in terms of $\langle \beta_i \rangle$. The closed-form expression obtained in Eq.\,\ref{qBfactor} reflects the fact that upon measuring a specimen's temperature one has access only to the temperature of a single $i$th domain, i.e., a region of the specimen. This may not necessarily hold in every case. For instance, considering voltage measurements over time during a capacitor discharge, to be discussed in Section\,\ref{applicationcapacitor}, we instead measure the voltage of the system as a whole rather than of a specific domain, so that the index $i$ is not required in such a case. The same holds true for measuring pressure decay over time when pumping a container, also to be discussed in Section\,\ref{applicationcapacitor}. The key point is that both are described macroscopically within the framework of hyperstatistics, since both cases probe the system as a whole, i.e., at the macroscopic level. A similar reasoning associated with temperature measurements can be extended to the analysis of brain activity, where electrical signals are recorded by distinct electrodes placed at different positions on the patient's head \cite{afewremarks}. The signal obtained by each electrode provides access only to local brain activity, i.e., at the domain level, analogously to a temperature measurement as discussed above.
Figure \ref{Fig-2} illustrates the quintessence of our approach. It is well-known that it is not meaningful to measure, for instance, the temperature of a single atom or particle. Instead, for a macroscopic system, the measured temperature of a certain domain is the result of the average of slightly distinct temperatures within it \cite{afewremarks}, which is given by $\langle \beta_i \rangle$ in Eq.\,\ref{qBfactor}. In other words, in our approach,  $\langle \beta_i \rangle$ refers to the expected value of 
$\beta_i$ of a given domain, whereas $\langle \beta \rangle$ is associated with the average of  $\langle \beta_i \rangle$ for the whole system, i.e., $\langle \beta \rangle =\ \langle\langle\beta_i\rangle\rangle$, cf.\,Figs.\,\ref{Fig-2} a), b), and c).
It is clear that $B_q^{n}$ (Eq.\,\ref{qBfactor}) incorporates organically the $q$-exp function, i.e., the distribution of Boltzmann factors in the domains is taken into account properly by $\Laplace\{\gamma\} = \exp_q(-\langle \beta_i \rangle E)$, cf.\,illustrated in Fig.\,\ref{Fig-2} b) and d). Putting in another way, by doing the Laplace transform of the $\gamma$-distribution function, having the Boltzmann factor as the kernel of the integral, we could say that we migrate from Boltzmann-Gibbsian to the phase-space of $q$-statistical mechanics. In other words, the $\Laplace\{\gamma(\beta_i)\}$ results in the expected value of $e^{-\beta_i E}$, i.e., $\langle e^{-\beta_i E} \rangle = \exp_q(-\langle \beta_i \rangle E) = \Laplace\{\gamma(\beta_i)\}$, cf.\,Ref.\,\cite{mahon}. Hence, it is evident that only the $\gamma$-distribution function gives rise to 
a mathematical structure that enables to move from Boltzmann-Gibbsian to the phase-space of $q$-statistical mechanics.
More specifically, the Boltzmann factor is extended, i.e., generalized in terms of the $q$-exp function, to take into account non-Boltzmann-Gibbsian  fluctuations in domains of the system accordingly. In the language of Bayesian statistics, the Boltzmann factor is the prior, while the generalized Boltzmann factor the posterior distribution of the data set of interest. Hence, Boltzmann factor works as a conjugated prior of the $q$-generalized Boltzmann factor \cite{Bayesian}.  As can be seen from Eq.\,\ref{qBfactor}, $B_q^{n}(E)$ assumes a $q$-exp form regardless of the PDF, cf.\,Table\,\ref{Table-1}. 
\begin{table*}[!t]
\centering
\begin{tabular}{@{}c@{}}
\includegraphics[width=\textwidth]{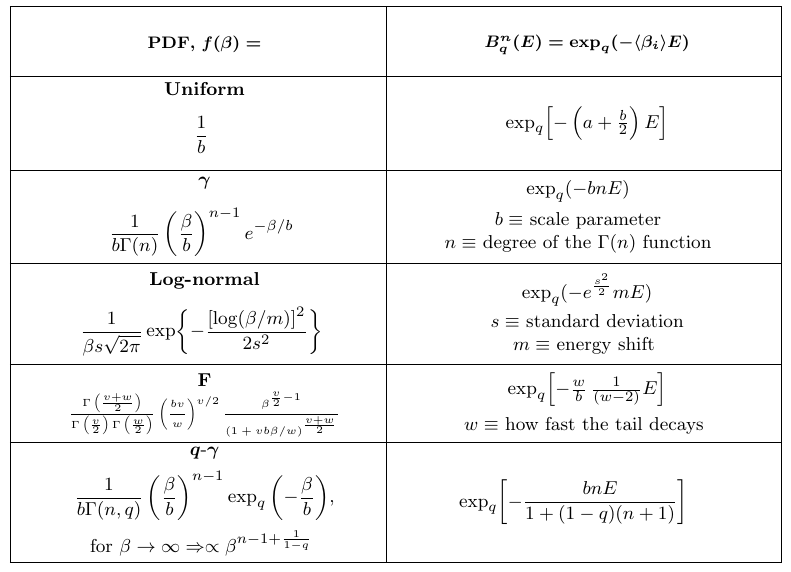}
\end{tabular}
\caption{\footnotesize \textbf{Hyperstatistics for various probability distribution functions at the domain (mesoscopic) level}. Mathematical closed forms of $B_q^{n}(E)$ obtained for distinct probability distribution functions, being the employed notation for the various PDFs based on Ref.\,\cite{beck2003}. The $q$-$\gamma$ PDF is first introduced in this work, see also Eq.\,\ref{q-gamma2}.}
\label{Table-1}
\end{table*}
This is because of the normalization condition required to all PDFs considered. Although the introduction of $B_q^{n}$ is required to reach such an important conclusion, i.e., all $B_q^{n}$s are given in terms of a $q$-exp function, the consideration of a $\gamma$-type distribution of Boltzmann factors suffices for the computation of $B_q^{n}$. Thus, the computation of the distribution of the $q$-exp functions associated with the domains of the system for any PDF becomes trivial, cf.\,Figs.\,\ref{Fig-2} b), c), and e), since it depends solely on $\langle\beta_i\rangle$, which indeed is PDF dependent (Supplementary Section 6). In other words, within each domain, the $q$-exp function arises only from the Laplace transform of the $\gamma$-distribution. Note that the computation of $B_q^n$ is PDF-independent, but $\langle\beta_i\rangle$ indeed depends on the chosen PDF. It turns out that analogously to the energy distribution of molecules in (monoatomic) polyatomic gases that (Maxwell) Boltzmann considered to derive the well-known Boltzmann factor \cite{maxwell1860,Boltzmann1868}, in our approach we consider a distribution of Boltzmann factors within a given domain (see Fig.\,\ref{Fig-2}) in order to derive $B_q^{n}(E)$. The use of the term \emph{domain} here was based on the seminal papers by Kolmogorov \cite{kolmogorov1, kolmogorov2}, being that such a term can be also seen as analogous to the ferromagnetic case \cite{apg}.
Remarkably, hyperstatistics obeys the principle of maximum entropy, i.e., the concavity of the entropy is preserved (Supplementary Section 1). Also, it is remarkable that for $q \rightarrow 1$ and making $\langle \beta_i \rangle = \beta$ the well-known Boltzmann factor is nicely recovered.
Table \ref{Table-1} summarizes the expressions obtained for $B_q^{n}(E)$ for the various PDFs considered by us in this work. Interestingly, all investigated cases of PDFs share the same mathematical structure, i.e., a $q$-exp function, with distinct arguments, i.e., $\langle \beta_i \rangle$, for each case. 
It is to be noted that one could consider distinct PDFs to analyze the distribution of Boltzmann factors within a given domain, a procedure at some extent reminiscent of superstatistics \cite{beck2003}. It is also worth emphasizing that analytical closed-form mathematical expressions for $B_q^{n}(E)$ were obtained for all PDFs here investigated, cf.\,Table \ref{Table-1}, being each of them normalizable (Supplementary Section 6). However, if one considers a given PDF to take into account the distribution of Boltzmann factors within a given domain, it turns out that by doing so the obtained $B_q^{n}$s assume either non closed-forms or more sophisticated mathematical forms, including, for instance, hypergeometric functions (Supplementary Section 7). In the case of hyperstatistics, any PDFs can be employed to compute $\langle \beta_i\rangle$ and thus to obtain $B_q^{n}$ in closed form, cf.\,Table \ref{Table-1}. 
Note that superstatistics considers BG-like fluctuations of $\beta$ in the cells composing the system \cite{beck2003}, whereas in our approach we consider non-Boltzmann-Gibbsian statistics for each domain of the system.  Also, note that $B_q^{n}$ is given naturally in terms of $\langle \beta_i \rangle$ and not solely on $\beta$. Essentially, $\langle \beta_i \rangle$ weights  the occupation of the energy levels. It is clear that one could consider a distribution of the energy $E$ instead of $\beta$ using the $\gamma$-distribution, which naturally leads to a $B_q^{n}(\beta)$ in terms of $\langle E_i \rangle$ (Supplementary Section 8). For the case of energy distribution, we are able to occupy the phase-space. In the classical framework of Maxwell–Boltzmann statistics, a distribution of velocities, i.e., kinetic energies, is considered, and the average energy of the system is obtained from the velocities distribution. Our proposal goes a step further. First, we consider a $\gamma$-distribution of Boltzmann factors at the level of domains, which leads to a $q$-generalized Boltzmann factor for each domain of the system, cf.\,Fig.\,\ref{Fig-2} b). Note that each Boltzmann factor inside every domain of the system is related to the probability of occurrence of a given $\beta_i$. Then, in a second step, a distribution of such $q$-generalized Boltzmann factors for all the domains of the system is considered, yielding a $B_q^{n}(E)$ for the system in terms of an average temperature, cf.\,Fig.\,\ref{Fig-2} f). The consideration of a given distribution of $q$-generalized Boltzmann factors regarding all domains within the system can thus be seen as the estimate of the ``probability of the probability''. In other words, we may say that hyperstatistics treats the $\gamma$-distribution of the canonical distribution \cite{kittelthermal}.
Computing the average of $kT_i$ rather than $\beta_i$ employing a $q$-$\gamma$-distribution, where $k$ is Boltzmann constant and $T_i$ the temperature of an $i$-th domain, leads to $\langle kT_i \rangle = \int_0^\infty (kT_i)\, q\text{-}\gamma(1/kT_i)\, d(1/kT_i) = (nkT_0)[1+n(1-q)]/(n-1)$, with $T_0$ the average temperature of a domain in the limit $q\to1$ and $n\to\infty$. It is worth recalling that $\langle kT_i\rangle \neq 1/\langle\beta_i\rangle$, since the average of an inverse differs from the inverse of the average whenever fluctuations are present \cite{thomas}. Note that the mathematical expression for $\langle kT_i \rangle$ is valid for $n > 1$ and lies in the convergence domain of $\Gamma(n, q)$, i.e., $1 < q < (1 + 1/n)$. Also, the mathematical form of $B(E) = \exp_q{(-\beta_0 E)}$, obtained in the frame of superstatistics with $q = (1 + 1/n)$ \cite{beck2003}, lies precisely on the boundary of this domain, cf.\,black solid line in Fig.\,\ref{Fig-1} b). Hyperstatistics in turn incorporates the whole range $1 < q < (1 + 1/n)$, making $q$ and $n$ independent parameters. Together with the consideration of a non-Boltzmann-Gibbsian statistics at the domain level, cf.\,details in Fig.\,\ref{Fig-2}, this is one of the main distinctions between hyperstatistics and superstatistics. More details on the distinctions between hyperstatistics and superstatistics will be discussed in Section\,\ref{distinguishing}. The canonical Boltzmann factor is recovered for $q \rightarrow 1$ in $B(E)$ and, if and only if, for $\lim_{n\to\infty}\lim_{q\to1} B_q^{n}(E)$. We stress that in the frame of superstatistics, a distribution of $\beta$ is considered \cite{beck2003}. In our approach, however, we have deduced a distribution of Boltzmann factors in terms of $\langle \beta_i \rangle$, being the latter indeed the observable accessible experimentally.
When compared with superstatistics, hyperstatistics is suitable for systems in which BG statistics already breaks down within individual domains of the system, while providing a ($q, n$)-generalized description that incorporates the average distributed quantity and an additional degree of freedom through $n$.
Essentially, our approach, based on $q$ and $n$, enables us to quantify both the deviation from the standard exponential decay through $q$ and the deviation of $\beta_{i}$ from its corresponding average, i.e., $\langle \beta_i \rangle$, through $n$. The latter in turn comes from the Mellin transform of the $q$-exp function, cf.\,Eqs.\,\ref{Mellintransform} and \ref{GammaMellin}. In other words, $q$ and $n$ can be interpreted as quantifiers of the degree of complexity in a given system, cf.\,Supplementary Sections 9-11. In particular, considering the role played by the parameter $n$ in the mathematical expression of the $\gamma$-distribution, larger values of $n$ may be associated with a less heterogeneous system. Essentially, lower values of $n$ lead to a broader decay of $B_q^n(E)$, which in turn is associated with an increased heterogeneity of the system in the sense that a higher number of accessible values of the observable of interest, e.g., relaxation time or pressure, is inherent to the system, being that BG statistics is recovered for $n \to \infty$. Interestingly, the product of $nq$, together with the analysis of both $q$ and $n$ individually suggests ``classes of universalities'', cf.\,discussions in the Supplementary Section 12. This is sustained when comparing $n =$ 4.032 for the discharge of a capacitor with $n =$ 4.553 for the midrapidity data set, cf.\,discussions in Section \ref{applications}, being that the case of the capacitor discharge corresponds to a more complex physical scenario. This is because in a real capacitor, the dielectric medium embodies intrinsic disorder, e.g., defects, while the particles produced in $p$-Pb collisions strongly interact and rapidly relax before reaching the detectors. For a given $q$-value, depending on the probability distribution function used, we have a set of families of $B_q^{n}$.
The proposal of the nonadditive $q$-entropy $S_q$ \cite{tsallis1988} generalized BG statistical mechanics, while in the approach of Beck and Cohen \cite{beck2003}, $q$-statistics appears as a particular case of superstatistics. However, we demonstrate here that the $q$-statistics framework dictates the behaviour of $B_q^{n}$, in the sense that, regardless of the PDF chosen to compute the distribution of the $q$-exp factor for all domains comprised in the system, $B_q^{n}$ is given by $B_q^{n}(E)=\exp_q(-\langle\beta_i\rangle E)$, with $\langle\beta_i\rangle$ depending on the specific PDF considered. Therefore, it is clear that our proposal of hyperstatistics is quite distinct from Beck and Cohen approach \cite{beck2003}. Although each PDF employed for computing $\langle \beta_i \rangle$ gives rise to distinct $B_q^{n}$, cf.\,Table \ref{Table-1}, it becomes evident that our approach is unique in the sense that it is dictated by $q$-exp-type functions with closed form having $n$ as a quantifier of heterogeneity. Thus, an advantage of using hyperstatistics refers to the fact that the specific distribution that works for the considered data sets is naturally embedded in $B_q^{n}$. In other words, upon employing $B_q^{n}$ we obtain an appropriate description of the data set regardless of the PDF considered as usual in other frameworks. Also, our approach should not be confused with the so-called hyperensembles \cite{crooks}, which considers distinct PDFs for the constituents of the system, named by Crooks as metaprobability. In contrast, hyperstatistics considers a unique PDF that describes the distribution of Boltzmann factors within a given domain, resulting in a ($q, n$)-generalized Boltzmann factor described by a $q$-exp function in terms of $\langle \beta_i \rangle$. It is worth mentioning that the hyperstatistics approach can be naturally applied to the case of a distribution of probabilities, cf.\,Eq.\,2.8 of Ref.\,\cite{beck2004}, yielding a $(q,n)$-Gaussian function in terms of $\langle \beta_i \rangle$. Following previous discussions, it becomes clear that superstatistics is based on BG equilibrium for each domain, while hyperstatistics considers them to follow non-Boltzmann-Gibbsian statistics. As a consequence, instead of assuming that fluctuations act on intensive quantities, hyperstatistics considers that fluctuations affect directly the statistical weights, i.e., we consider a distribution of Boltzmann factors in the domains composing the system. This is totally in line with the fact that the presence of the so-called multiplicative noise leads to nonextensive statistical mechanics \cite{celia2003}. Also, the auto-correlation function plays a key role in the context of hyperstatistics, since it quantifies memory effects within a given system, which are in turn associated with the long-tailed decays discussed in the present work. In this regard, as an additional direct application of hyperstatistics, we refer to the Griffiths-like phase \cite{griorigin,GR1,GR2}. The latter is characterized by a long-tail exponential decay of the auto-correlation function \cite{gri1}, which is captured by magnetization measurements as a function of time in the vicinity of the transition temperature, see, e.g., Ref.\,\cite{gri2}. Yet, hyperstatistics is in line with the observation of non-Boltzmann-Gibbsian statistics right at critical points \cite{prbl,facets}, being that the block size may be seen as analogous to the domains composing the system.

\subsection{Macroscopic level}\label{macroscopicBH}

As a next step, following previous discussions regarding the relevance of considering non-Boltzmann-Gibbsian behaviour at the domain level, in what follows we propose the hyperstatistical ($H$) generalized Boltzmann factor $B_H(E)$ valid at the macroscopic level, namely:
\begin{equation}
  B_H(E)= \langle \exp_q(-\langle \beta_i \rangle E) \rangle = \int_0^{\infty} q\text{-}\gamma(\langle \beta_i \rangle)\,\exp_q(-\langle\beta_i \rangle E)\,d\langle\beta_i\rangle =  \Laplace_q\{q\textmd{-}\gamma(\langle\beta_i\rangle)\},
  \label{BH}
\end{equation}
where $\Laplace_q\{q\textmd{-}\gamma(\langle\beta_i\rangle)\}$ refers to the $q$-generalized Laplace transform of the $q$-$\gamma$ PDF and the $i$ index refers to the label of the considered domain, cf.\,Fig.\,\ref{Fig-2}, being the $q\text{-}\gamma(\langle \beta_i \rangle)$ given by (Table\,\ref{Table-1}):
\begin{equation}
  q\text{-}\gamma(\langle\beta_i\rangle)=\frac{1}{b^*\Gamma(n, q)}\left(\frac{\langle\beta_i\rangle}{b^*}\right)^{n-1}\exp_q\!\left(-\frac{\langle\beta_i\rangle}{b^*}\right),
  \label{q-gamma2}
\end{equation}
where $b^* = \langle\beta\rangle/n$. The use of a $q$-$\gamma$ distribution in Eq.\,\ref{BH} is motivated by the hierarchical structure inherent to hyperstatistics, in which the domains at the mesoscopic level are described by $q$-exp functions. In analogy with the conventional $\gamma$-distribution, which arises from a Poisson process involving sums of exponentials, cf.\,Supplementary Section 3, the construction of the macroscopic behavior requires an analogous statistical treatment. In such a case, however, the relevant building blocks are $q$-exp functions rather than ordinary exponential functions. As a consequence, the corresponding distribution at the macroscopic level is a $q$-$\gamma$ distribution, which is then employed in the determination of $B_H(E)$. Thus, plugging  Eq.\,\ref{q-gamma2} into Eq.\,\ref{BH} and rearranging, we have:
\begin{align}
  B_H(E)&=\frac{1}{b^*\,\Gamma(n, q)}\int_0^{\infty}\left(\frac{\langle\beta_i\rangle}{b^*}\right)^{n-1}\exp_q\!\left(-\frac{\langle\beta_i\rangle}{b^*}\right)\exp_q(-\langle\beta_i\rangle E)\,d\langle\beta_i\rangle.
\end{align}
Defining $m = 1/(q-1)$, $c = b^*E$, and $\phi = (q-1)\langle\beta_i\rangle/b^*$, $B_H(E)$ can be written as:
\begin{equation}
B_H(E) = \frac{1}{\Gamma(n,q)\,(q-1)^n} \int_0^{\infty} {\phi}^{n-1} (1+\phi)^{-m} (1+c\phi)^{-m}\, d\phi .
\end{equation}
Now, making another change of variables $s = \phi/(1+\phi)$, $B_H(E)$ now reads:
\begin{equation}
B_H(E) = \frac{1}{\Gamma(n,q)\,(q-1)^n} \int_0^{1} s^{n-1} (1-s)^{2m-n-1} \left[1-(1-b^*E)s\right]^{-m} ds .
\label{int1}
\end{equation}
The integral in Eq.\,\ref{int1} can be evaluated in closed form, yielding:
\begin{equation}
B_H(E) = \frac{1}{\Gamma(n,q)\,(q-1)^n}\, \Gamma(2m-n)\,\Gamma(n)\; {}_2\tilde{F}_1(m,n;2m;1-b^*E),
\end{equation}
where ${}_2\tilde{F}_1$ is the regularized Gauss hypergeometric function \cite{gradsteyn}. Note that the hypergeometric function $_2\tilde{F}_1$ emerges naturally as the exact $q$-generalized Laplace transform of the $q$-$\gamma$ distribution, with the $q$-exp acting as the statistical kernel of the transform. It therefore encodes the combined effects of microscale, i.e., local-scale fluctuations of the physical parameter, e.g., $\beta_i$ or $\tau_i$, and the intrinsic non-Boltzmann-Gibbsian character of the statistical weight at the domain level. Employing Eq.\,\ref{gammabeta} and making use of the relation between the Beta and $\Gamma(n)$ functions, $B_H(E)$ reads:
\begin{equation}
B_H(E) = \frac{\Gamma(2m-n)\,\Gamma(m)}{\Gamma(m-n)} \cdot {}_2\tilde{F}_1(m,n;2m;1-b^*E).
\end{equation}
The regularized hypergeometric function is related to the standard one by \cite{gradsteyn}:
\begin{equation}
{}_2\tilde{F}_1(m,n;2m;1-b^*E) = \frac{{}_2F_1(m,n;2m;1-b^*E)}{\Gamma(2m)},
\end{equation}
so that by using an identity that relates hypergeometric and $\Gamma$ functions \cite{gradsteyn}, we obtain:
\begin{equation}
\boxed{\,B_H(E) = \frac{{}_2F_1(m, n; 2m; 1 - b^*E)}{{}_2F_1(m, n; 2m; 1)}\,,}\hspace{1.0cm} B_H(0) = 1,
\label{B_H(E)}
\end{equation}\\
being ${}_2F_1(m,n;2m;z) = \sum_{k=0}^{\infty} \frac{(m)_k\,(n)_k}{(2m)_k\, k!}\, z^k$ with $(m)_k$, $(n)_k$, and $(2m)_k$ denote the Pochhammer symbols \cite{gradsteyn}. We now analyze the limit $m = \frac{1}{(q-1)} \to \infty$, which corresponds to $q \to 1^{+}$. Taking the limit of each factor separately, namely:
\begin{equation}
\lim_{m\to\infty} B_H(E) = \underbrace{\lim_{m\to\infty} \frac{\Gamma(2m-n)\,\Gamma(m)}{\Gamma(m-n)\,\Gamma(2m)}}_{2^{-n}} \cdot \underbrace{\lim_{m\to\infty} {}_2F_1(m,n;2m;1-b^*E)}_{2^{n}(1+b^*E)^{-n}} ,
\end{equation}
which gives:
\begin{equation}
\lim_{m\to\infty} B_H(E) = 2^{-n} \cdot 2^{n} (1+b^*E)^{-n} .
\end{equation}
Therefore,
\begin{equation}
\boxed{\lim_{m\to\infty} B_H(E) = (1+b^*E)^{-n} = B(E),}
\end{equation}\\
i.e., the functional form of the superstatistical Boltzmann factor for the $\gamma$ PDF is recovered, but with $\langle \beta \rangle = \langle \beta_0 \rangle$ instead of $\beta_0$, since hyperstatistics has three hierarchical levels while superstatistics has two. Hence, in the limit $q \to 1^+$, the domains recover BG statistics, as assumed in the superstatistics framework, while for $q > 1$, their non-BG character is explicitly taken into account for a macroscopic description of the system. Thus, Eq.\,\ref{B_H(E)} captures how deviations from BG statistics at the domain level influence the macroscopic behavior of the system, naturally recovering superstatistics in the limit $q \to 1^+$. Also, upon computing the limit of $n \rightarrow \infty$, the canonical Boltzmann exponential weight is recovered. Hence, our proposal of $B_H(E)$ naturally incorporates both superstatistics for the case of the $\gamma$ PDF and BG formalisms. In its quintessence, hyperstatistics considers that the system is governed by BG statistics at the microscopic level and by $q$-statistics at the mesoscopic one, i.e., at the domain level, cf.\,Fig.\,\ref{Fig-2}. Hence, $B_H(E)$ should be regarded as the expected value of $\exp_q(-\langle\beta_i\rangle E)$, i.e., $\langle \exp_q(-\langle \beta_i \rangle E)\rangle$, cf.\,Eq.\,\ref{BH}, at the macroscopic level (Fig.\,\ref{Fig-2}). Note that in hyperstatistics, the Boltzmann factor $\exp(-\beta E)$ is assumed to be valid at the microscopic level. Its expected value within a domain in terms of $\Laplace\{\gamma(\beta_i)\}$ gives rise to $B_q^n(E)$ in terms of a $q$-exp function, which describes the mesoscopic (domain) level. At the next hierarchical level, $B_H(E)$ is obtained by computing the expected value of the domain-level factors $\exp_q(-\langle \beta_i \rangle E)$ over the whole system. Hence, the hierarchy of hyperstatistics can be summarized as $\exp(-\beta E)$ (microscopic), $B_q^n(E)$ (mesoscopic/domain), and $B_H(E)$ (macroscopic). It is thus essential, when applying hyperstatistics to fit experimental data, to distinguish whether the measurements are associated with either the mesoscopic (local, i.e., domain level), or macroscopic level of the system. In the following, we demonstrate the applicability of hyperstatistics at both levels.

\section{Discussions and applications of hyperstatistics}\label{applications}
In what follows, we demonstrate the applicability of hyperstatistics by presenting experimental results for the discharge of a capacitor considering a $\gamma$-type distribution of relaxation times; the pressure decay over time upon pumping $^4$He lines of a cryostat; fits of the midrapidity data for \emph{p}-Pb collisions at the LHC \cite{alice}; and acceleration in turbulent systems,  as well as the deduction of the \emph{universal} power-law dielectric response of non-homogeneous systems. Astonishingly, our approach accurately captures the evolvement of all these experimental data set. It is worth emphasizing that Beck and Cohen in their original paper \cite{beck2003} considered a distribution of intensive parameters, e.g., temperatures \cite{prl07}, whereas here we apply our proposal of hyperstatistics at the level of local dynamics, e.g., relaxation times. Our approach, based on a ($q, n$)-generalized Boltzmann factor, enables us to explore with accuracy  the power-law-type asymptotic behaviour embedded in various distinctly physical systems, cf.\,applications below. Throughout this Section, the quality of the fits is quantified by the coefficient of determination $R^2 = 1 - [\sum_i (y_i - f_i)^2]/[\sum_i (y_i - \bar{y})^2]$, where $y_i$ is the $i$th experimental value, $f_i$ is the corresponding value predicted by the fitting function, and $\bar{y}$ is the mean value of the experimental data. The mathematical expression used to determine $R^2$ is implemented directly in the fitting codes employed for all analyzed data sets, being all codes available in Ref.\,\cite{zenodo}. Thus, $R^2$ quantifies how well the experimental data are described by the fitting function, i.e., the closer $R^2$ is to unity, the more accurately the fitting function reproduces the experimental data \cite{hughes,young}. The obtained $R^2$ and mean values, as well as the uncertainties in the fitting parameters obtained for all analyzed data sets, are summarized in Section 12 of the Supplementary Material.

\subsection{Series RC circuit and pumping of the $^4$He lines of a cryostat}\label{applicationcapacitor}
In an ideal series RC circuit, the discharge of a capacitor is governed by a simple linear first-order differential equation, leading to a purely exponential relaxation \cite{tipler}:
\begin{equation}
V(t) = V_0 \cdot e^{-t/\tau},\label{capacitorbasico}
\end{equation}
where $V$ is the voltage decay, $t$ time, $V_0$ the initial voltage,  and $\tau = RC$ the relaxation time, being $R$ and $C$ the nominal values of both the electrical resistance and capacitance, respectively. This well-known behaviour reflects the existence of a single characteristic $\tau$ and the absence of memory effects. However, real capacitors are built with dielectric media, such as polymers, glasses, or composite materials  with intrinsic disorder \cite{composite}. In such systems, subtle local fluctuations of both the electrical resistance of the dielectric and capacitance give rise to a broad distribution of relaxation times, a behaviour quite distinct from that expected for textbook capacitors \cite{tipler}. As a consequence, $V(t)$ deviates from the ideal pure exponential-like charge or discharge curves. Instead, $V(t)$ can be described by stretched-exponential or power-law-type behaviour, as broadly reported in dielectric response experiments \cite{cole1941, davidson1951, havriliak1967, jonscher1977, kremer2003}. 
\begin{figure}[!h]
\centering
\includegraphics[width=0.7\textwidth]{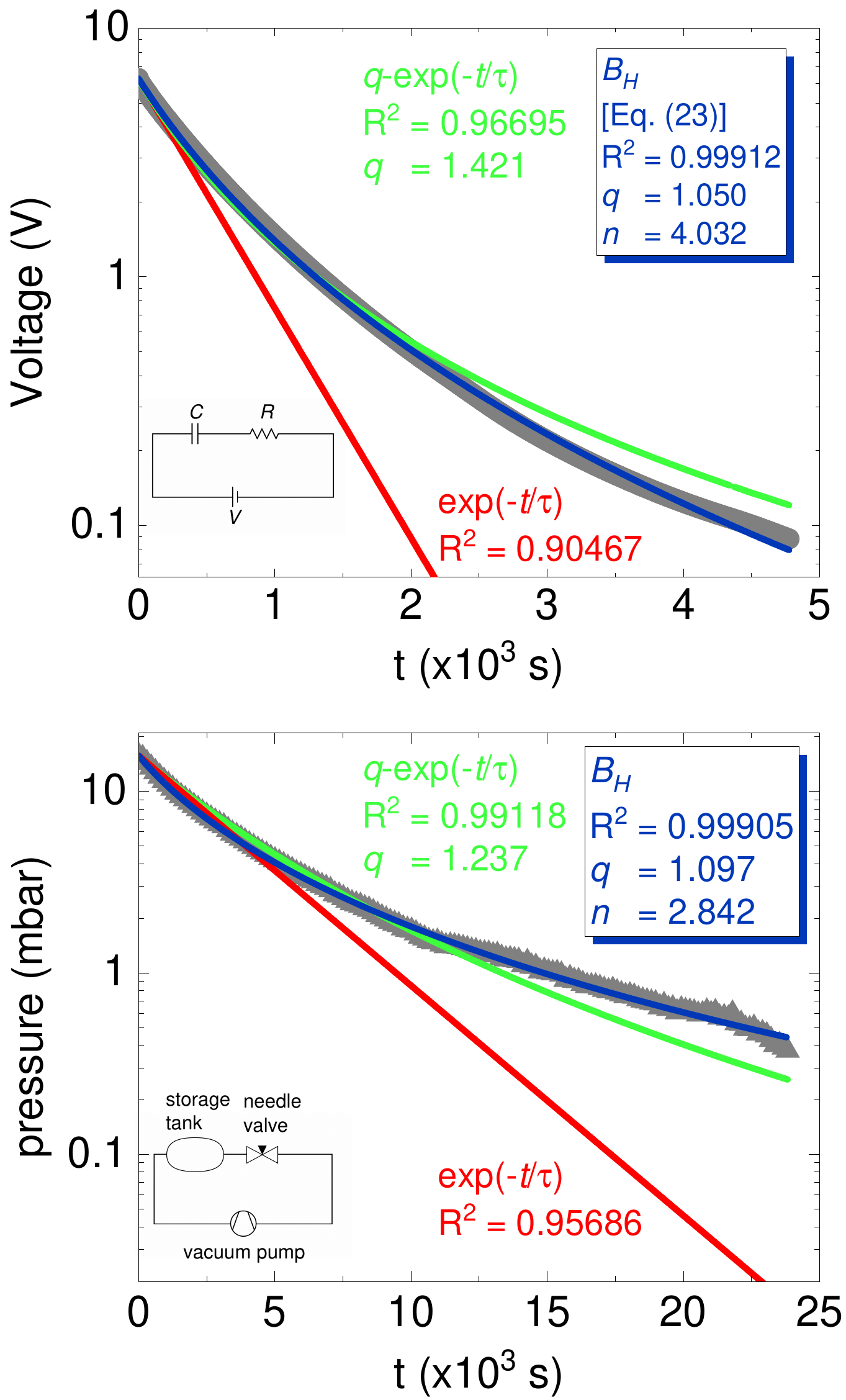}
\caption{\footnotesize \textbf{The discharge of a capacitor in a table top $RC$ circuit and pumping of $^4$He lines of a cryostat}. Top: Voltage \textit{versus} time, i.e., $V(t)$, during the discharge of the capacitor. Bottom: Pressure \textit{versus} time, i.e., $p(t)$, regarding the pressure decay during the pumping of the $^{4}\mathrm{He}$ circulation lines in a Teslatron-PT cryostat. The data sets shown in both panels were fitted using the canonical exponential, $q$-exponential function, and hyperstatistics at the macroscopic level, i.e., $B_H$ (Eq.\,\ref{B_H(E)}) with $b^{*}E = t/(n\langle\tau\rangle)$. The fittings for the pressure \emph{versus} time data set were made employing $p(t) = p_0\exp{(-t/\tau)}$ (red line), $p(t) = p_0\exp_q{(-t/\tau)}$ (green line), and $B_H$ (Eq.\,\ref{B_H(E)}) (blue line), where $p_0$ is the initial pressure. The insets of both panels depict the circuits regarding the setup to measure the discharge of the capacitor and the pumping of the $^{4}\mathrm{He}$ circulation lines in a Teslatron-PT cryostat. The corresponding values of R$^2$, $q$, and $n$ for each fit are depicted accordingly.}
\label{Fig-3}
\end{figure}
In this context, an appropriate physical description consists in considering that $\tau$ itself is $\gamma$-distributed ($\chi^2$). At this point we recall Eq.\,6 of Ref.\,\cite{wilk2000}, more details in Section\,3 of the Supplementary Material, namely:
\begin{equation}
f(1/\lambda) = \frac{1}{\Gamma(n)}(n\lambda_0)\left(\frac{n\lambda_0}{\lambda}\right)^{n-1}\exp{\left(-\frac{n\lambda_0}{\lambda}\right)},
\label{prl}
\end{equation}
where $\lambda$ is a distributed parameter and $\lambda_0$ its average. In our case, we recognize $\lambda = \tau$ and $\lambda_0 = \langle \tau \rangle$, so that by rearranging Eq.\,\ref{prl}, we have:
\begin{equation}
f(1/\tau) = \frac{1}{\Gamma(n)}(n\langle \tau \rangle)^n \tau^{1-n}\exp{\left(-\frac{n\langle \tau \rangle}{\tau}\right)},
\label{1/tau2}
\end{equation}
where $\langle \tau \rangle$ fixes the characteristic relaxation scale of the distribution of $\tau$, being the normalization condition $\int_0^\infty f(1/\tau)\, d(1/\tau) = 1$ fulfilled.
The key point here is that our consideration of $f(1/\tau)$  for the relaxation times is not by chance. Instead, its choice was based on the fact that the Laplace transform of the $\gamma$-distribution leads to a $q$-exp, cf.\,discussions in Section \ref{proposal}. In the discharge regime, the relaxation time becomes broadly distributed due to the inherent imperfections of a real capacitor, so that the $\gamma$-distribution is appropriate to describe the system.
In this framework, $V(t)$ corresponds to a superposition of exponential-like relaxation times, i.e.:
\begin{equation}
\label{Vt}
V(t) = V_0 \int_0^\infty e^{-t/\tau} f(1/\tau)\, d(1/\tau).
\end{equation}
Note that the argument of the exponential function in Eq.\,\ref{Vt} embodies $1/\tau$. The latter is distinct from the distribution of $\beta_i$ associated with an exponential function incorporating $\beta_i$ in its argument, cf.\,Eq.\,\ref{Bfactor}. Also, note that the sub-index $i$ is omitted in $\tau$ due to the fact that we are dealing with a physical quantity that mirrors the behaviour of the system as a whole, and not locally, i.e., at the domain level, as is the case of $\beta_i$. Plugging Eq.\,\ref{1/tau2} into Eq.\,\ref{Vt}, we achieve:
\begin{eqnarray}
V(t) &=& \frac{V_0 (n \langle \tau \rangle)^n}{\Gamma(n)} \int_0^\infty \tau^{1-n}\exp{\left[-\frac{(n \langle \tau \rangle + t)}{\tau}\right]} d(1/\tau).
\end{eqnarray}
Solving the integral and making some simplifications, we have:
\begin{eqnarray}
V(t) &=& V_0 \left[\frac{n \langle \tau \rangle}{(n \langle \tau \rangle + t)}\right]^n.
\label{V(t)}
\end{eqnarray}
Now, rearranging the terms and simplifying Eq.\,\ref{V(t)}:
\begin{eqnarray}
V(t) &=& V_0 \left(1 + \frac{t}{n\langle \tau \rangle}\right)^{-n}.
\label{V(t)2}
\end{eqnarray}
By substituting $n = 1/(q-1)$ in Eq.\,\ref{V(t)2}, we achieve:
\begin{eqnarray}
V(t) &=& V_0\left[1 + (q-1)\frac{t}{\langle \tau \rangle}\right]^{\frac{1}{(1-q)}}.
\end{eqnarray}
The latter can be naturally recognized as a $q$-exp-type decay, cf.\,previous discussions, namely:
\begin{equation}
\label{Vtq}
V(t) = V_0 \cdot \exp_q\!\left(-\frac{t}{\langle \tau \rangle}  \right) = V_0 \cdot \exp_q\!\left[-\frac{(n-1)}{(n\tau_0)[1+n(1-q)]}t\right],
\end{equation}
with $\langle \tau \rangle
= \int_{0}^{\infty} \tau q\textmd{-}\gamma(1/\tau)\,d(1/\tau)
= \int_{0}^{\infty} \tau \frac{1}{\Gamma(n,q)}
\left(n\tau_0\right)^n
\tau^{1-n}
\exp_{q}\left(-\frac{n\tau_0}{\tau}\right)\,d(1/\tau)\\
= (n\tau_0)[1+n(1-q)]/(n-1)$ 
(Supplementary Section 5) the average relaxation time, being $\tau_0$ the average relaxation time in the limit of both $q \to 1$ and $n \to \infty$. In our analysis, we determine $\tau_0$ as the value in the abscissa at which the physical quantity of interest reaches 37\% of its initial value, i.e., the usual procedure in the analysis of the discharge of a capacitor \cite{tipler}. Such a protocol was then extended to all data fits shown in Figs.\,\ref{Fig-3}, \ref{Fig-4}, and \ref{Fig-5}. The adoption of this protocol for determining $\tau_0$ establishes a direct connection between the hyperstatistics approach and experimental physics, by tying the average of the distributed quantity to an experimentally defined observable. The consideration of a distribution of relaxation times is based on physical arguments, cf.\,previous discussions, and led to a refined mathematical expression for $V(t)$ (Eq.\,\ref{Vtq}). Hyperstatistics provides thus a more realistic behaviour not only of the discharge of a capacitor, but also to other ordinary exponential-decay processes cf.\,following discussions. 
Hence, our approach provides a clear cut physical criterion for the breakdown of descriptions based on single $\tau$ in non-ideal RC circuits and related phenomena. To demonstrate our proposal, we have used a table top RC circuit to measure the discharge of a capacitor, fitting the resulting data set with Eq.\,\ref{B_H(E)}, as well as employing both regular and $q$-exp functions for the sake of comparison, cf.\,top panel of Fig.\,\ref{Fig-3}. Note that, based on the discussions presented in Section\,\ref{proposal}, Eq.\,\ref{Vtq} is valid at the domain level. Since the voltage decay of a capacitor is a macroscopic measurement, we have employed $B_H$ to properly fit the data set of the capacitor discharge, as shown in the top panel of Fig.\,\ref{Fig-3}. The experimental data set shown in the top panel of Fig.\,\ref{Fig-3} makes it clear that $B_H$, i.e., hyperstatistics at the macroscopic level, provides a better description of the time dependence of the capacitor voltage in the discharging regime when compared to the usual $q$-exp based fitting.  It is worth mentioning that our approach is not phenomenological, since we are considering a distribution of relaxation times. Thus, upon employing a $\gamma$-distribution of $\tau$, the accuracy of our fitting procedure is improved. Also, we stress that for our case of study, i.e., standard series RC circuit, the step from Eq.\,\ref{Vt} to Eq.\,\ref{Vtq} is analytically exact when fluctuations are considered to follow a $\gamma$-distribution. To be more precise, the Laplace transform of $f(1/\tau)$, i.e.,  $V(t) = V_0 \cdot [\Laplace\{f(1/\tau)\}]$, can be evaluated exactly \cite{ricieri}. 

Another experimental case of study explored by us refers to the temporal pressure decay that occurs during the pumping of $^4$He lines of a cryostat. Following discussions in Ref.\,\cite{bello}, the pressure decay is usually considered to follow $p = p_0 \cdot e^{-t/\tau_p}$, where $p_0$ is the initial pressure and $\tau_p$ the ratio between the volume of the vessel and the pumping speed.  Upon analyzing the bottom panel of Fig.\,\ref{Fig-3}, it becomes evident that the data set is better fitted using hyperstatistics. This is because in a real pumping process, the pumped particles have a broad distribution of velocities. 
As a matter of fact, since the system is kept at room-temperature, gas particles already exhibit a distribution of thermal velocities. Also, there should be different flow regimes along the pumping line. Yet, particle-wall and particle-particle interaction cannot be neglected. As a consequence, an idealized situation in which all particles are removed from the vessel with the same velocity is not quite accurate. Instead, it is appropriate to consider a distribution of velocities of the gas particles being pumped. 
Hence, based on such physical arguments, a distribution of effective escape velocities can be naturally inferred, which makes our proposal to properly describe the behaviour of the pressure \emph{versus} time, cf.\,bottom panel of Fig.\,\ref{Fig-3}. At this point, an analogy can be made between the series RC circuit and the pumping of the $^4$He circulation lines attached to the cryostat, cf.\,insets in both top and bottom panels of Fig.\,\ref{Fig-3}. The storage tank retains gas particles analogously to the capacitor that stores electrical charges. The needle valve can be seen as the analogous to the resistor since it hinders the gas flow through the system, so that the flow of electrons in the RC circuit upon discharging the capacitor is analogous to the flow of gas particles during the pumping of the $^4$He lines. Note that for the pressure decay experimental data set we obtain $q =$ 1.097, while for the capacitor discharge process $q =$ 1.050, suggesting that the pressure decays more slowly during the pumping process than the voltage during the discharge of the capacitor. This difference mirrors the quite distinct physical mechanisms involved, i.e., while the capacitor begins to discharge ``almost immediately'' after the short circuit is made, the pumping of the $^4$He lines of the cryostat constitutes a slower process. The latter is related to the inherent difficulties posed to remove the broaden distributed $^4$He molecules plus moisture in the $^4$He lines by using a turbo vacuum pump. This is clearly seen in the time axes of the upper and lower panels of Fig.\,\ref{Fig-3}, where the pressure decay occurs over a time scale two orders of magnitude longer than that of the capacitor.

\subsection{Hyperstatistics and high-energy Physics}
\begin{figure}[!t]
\centering
\includegraphics[width=0.9\textwidth]{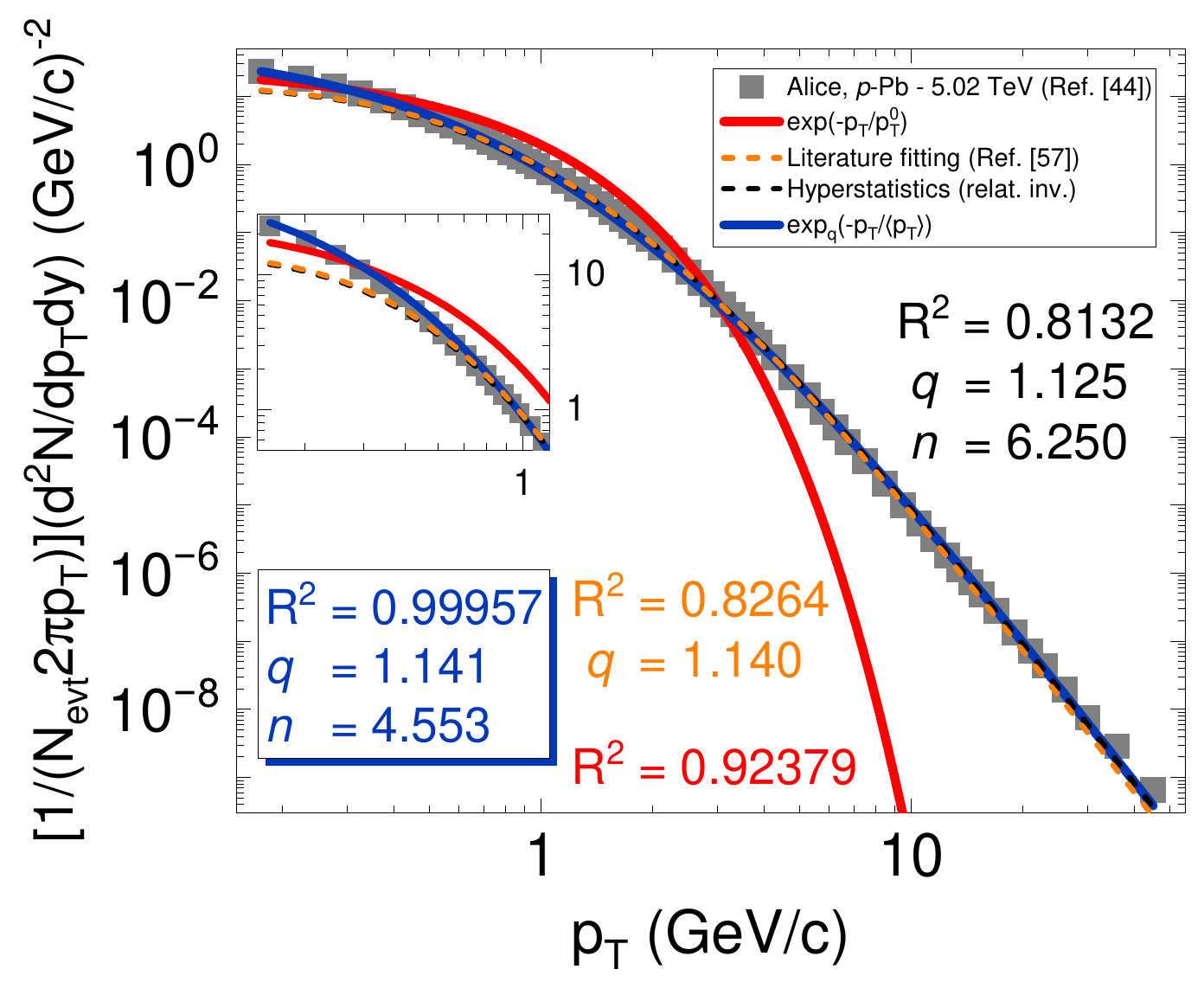}
\caption{\footnotesize \textbf{Midrapidity and hyperstatistics}. Transverse momentum $p_T$ distribution in $p$-Pb collisions considering a pseudorapidity $|\eta_{\textmd{cms}}| <$ 0.3. The data set was extracted from Ref.\,\cite{alice} and fitted with an exponential decay function $[1/(N_{evt}2\pi p_T)] d^2N/(dp_Tdy) = \nu\exp{(-p_T/p_T^0)}$ (red line), where $\nu$ is the highest value of $[1/(N_{evt}2\pi p_T)] d^2N/(dp_Tdy)$ associated with the data set and $p_T^0$ a characteristic transverse momentum, Eq.\,17 of Ref.\,\cite{cleymans} (orange color dashed line), a relativistic invariant approach of hyperstatistics based on Refs.\,\cite{cleymans,parvan2023} (black dashed line), and hyperstatistics, namely $[1/(N_{evt}2\pi p_T)] d^2N/(dp_Tdy) = \nu\exp_q{(-p_T/\langle p_T \rangle)} = 
\nu\exp_q{\{-p_T(n-1)/[(np_T^*)(1+n(1-q))]\}}$ (blue line), where where $p_T^*$ is the transverse momentum at which the first data point reaches 37\% of its first value, which coincides with $\langle p_T \rangle$ only for $q \to 1$ and $n \to \infty$. The values of R$^2$ for each fitting are shown in the corresponding colors, as well as the values of $q$ and $n$. The inset depicts the fitting of the data for low values of $p_T$.}
\label{Fig-4}
\end{figure}

Furthermore, we have applied our hyperstatistics approach to experiments in the field of high-energy Physics. The data set shown in Fig.\,\ref{Fig-4} depicts the transverse momentum $p_T$ distribution for $p$-Pb collisions in the frame of the
ALICE collaboration \cite{alice}.
It is remarkable that the most accurate fit to this data set is obtained by using our approach in terms
of hyperstatistics. This is particularly true because of the intrinsic $p_T$ distribution associated with the $p$-Pb collisions, which is nicely captured by hyperstatistics. Note that fitting the data set using our hyperstatistics framework properly captures the behaviour of the data set over the entire $p_T$ measured range, providing thus a more accurate description of the experimental data. Remarkably, the value of $q =$ 1.141 obtained in the frame of our proposal of hyperstatistics is in line with that reported in Ref.\,\cite{Deppman2020} regarding the Yang-Mills field theory for the $pp$ collision. Yet, we emphasize that even though the low $p_T$ region is governed by thermal effects while the high $p_T$ one is not \cite{rostovtsev}, the resulting transverse momentum distribution in the whole range can be effectively described by hyperstatistics. Note that a detailed derivation of the transverse momentum distribution \cite{parvan2023} is not required, since its mathematical structure is the same to that derived at the domain level for the voltage decay during the discharge of a capacitor. Hence, the mathematical procedure which leads to the fitting function obtained in the frame of hyperstatistics for the $p_T$ distribution, namely $\left[1/\left(N_{evt}2\pi p_T\right)\right]d^2N/\left(dp_Tdy\right)=\nu\exp_q(-p_T/\langle p_T\rangle)$, cf.\,caption of Fig.\,\ref{Fig-4}, follows the same mathematical steps as those used to obtain $V(t) = V_0\exp_q(-t/\langle \tau \rangle)$, cf.\,Eq.\,\ref{Vtq}. This is because usually $\left[1/\left(N_{evt}2\pi p_T\right)\right]d^2N/\left(dp_Tdy\right)$ \emph{versus} $p_T$ data is fitted with an exponential decay function, cf.\,Eq.\,(9) of Ref.\,\cite{adare2011}. The fit depicted by the orange color dashed line in Fig.\,\ref{Fig-4} represents the relativistically invariant Tsallis distribution, i.e., Eq.\,(17) of Ref.\,\cite{cleymans} and Eq.\,(48) of Ref.\,\cite{parvan2023}. Such a distribution was evaluated using the parameters reported in Ref.\,\cite{cleymans} for the same pseudorapidity interval $|\eta_{\mathrm{cms}}|<0.3$, namely $q=1.140$, $T=112.81$\,MeV, and $R=4.03$\,fm. Note that the fitting of the midrapidity data employing an adapted version of the relativistically invariant approach of hyperstatistics (black dashed line) does not yield a proper fitting, especially in the low- and high-$p_T$ regimes. In contrast, the genuine hyperstatistics approach (blue solid line) achieves a higher $R^2$ and properly describes both the low and high ranges of $p_T$ when compared with other approaches. Hence, rather than focusing on whether a relativistically invariant formulation is employed, our goal is to demonstrate the applicability and versatility of the hyperstatistics approach.

\subsection{Hyperstatistics and turbulence: Bodenschatz experiment}
\begin{figure}[!t]
\centerline{\includegraphics[width=0.93\textwidth]{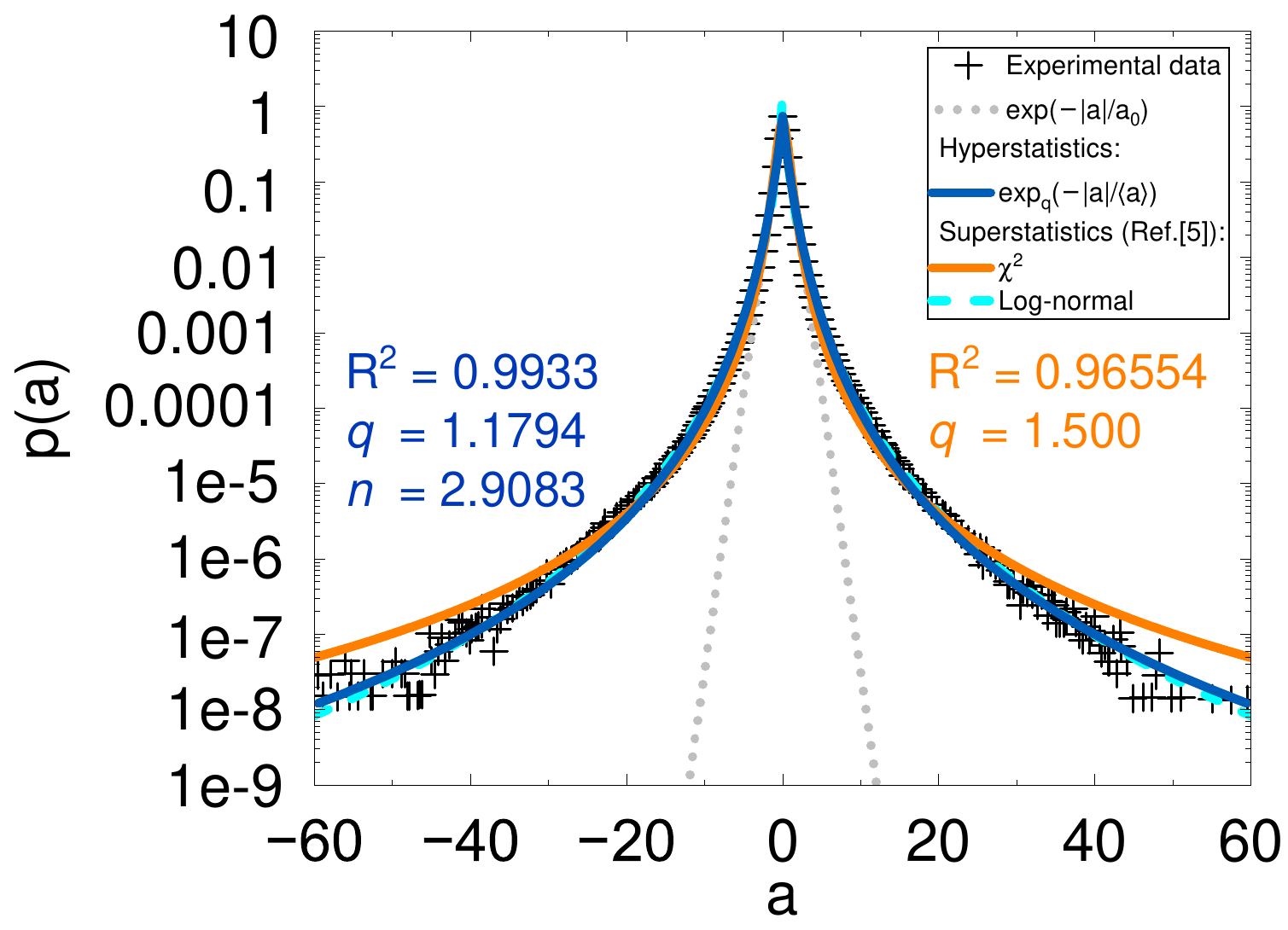}}
\caption{\textbf{Distribution of accelerations in turbulent systems.} Probability density functions of the transverse acceleration \textit{versus} acceleration normalized by its standard deviation \cite{bodenschatz}. The blue solid line refers to the fit employing hyperstatistics, i.e., $p(a) = \exp_q(-|a|/ \langle a \rangle) \,= \exp_q{\{-|a|(n-1)/[(na_0)(1+n(1-q))]\}}$, being the values of both $q$ and $n$ for other Reynolds numbers \cite{bodenschatz} shown in Section\,12 of the Supplementary Material. The orange solid line is related with the literature fitting employing superstatistics in terms of $\chi^2$-distribution, while the cyan dashed line refers to the case of a log-normal distribution, cf.\,Ref.\,\cite{beck2004}. Gray dotted line refers to a fit using a standard exponential function.}
\label{Fig-5}
\end{figure}

Hyperstatistics can be straightforwardly applied to the acceleration of a fluid particle in a turbulent flow in the frame of the Bodenschatz experiment \cite{bodenschatz}. Essentially, the acceleration of a fluid particle in turbulent flows arises from pressure gradients \cite{bodenschatz}, being properly described within the framework of Kolmogorov’s scaling theory \cite{kolmogorov1,kolmogorov2}. In simple words, in such a theory a domain is considered homogeneous when the PDF does not depend on spatial coordinates, time, or velocity \cite{kolmogorov1,kolmogorov2}. In our approach, based on Fig.\,\ref{Fig-2}, we consider a non-homogeneous domain characterized by a distribution of fluid particle accelerations.
Figure\,\ref{Fig-5} presents experimental data extracted from Ref.\,\cite{bodenschatz} showing the PDF \emph{versus} the transverse acceleration normalized by its standard deviation for distinct Reynolds numbers.
Note that, within the framework of hyperstatistics, the turbulence data shown in Fig.\,\ref{Fig-5} can be precisely fitted using a $\exp_q{(-\langle \beta_i \rangle E)}$, cf.\,Table\,\ref{Table-1}. In contrast, within the superstatistics framework, a proper fit of the data set depicted in Fig.\,\ref{Fig-5} requires the choice of the underlying PDF, cf.\,Fig.\,3.2 in Ref.\,\cite{beck2004}. This is because the author in Ref.\,\cite{beck2004} employed the superstatistics approach to consider a distribution of probabilities. In the frame of our approach, however, the distribution of Boltzmann factors in the scale of domains already incorporates a distribution of probabilities.  The proper use of the hyperstatistics formalism to turbulent systems, together with the other applications discussed in previous sections, highlights that the hyperstatistics framework can be applied to a wide variety of physical systems. 
Yet, using $B_q^{n}(E)$, we can analyze how the value of $q$ is affected by distinct turbulence data sets corresponding to different Reynolds numbers (Supplementary Section 9). In terms of the so-called Taylor microscale Reynolds number $R_{\lambda}$, the range 140 $<$ $R_{\lambda}$ $<$ 970 corresponds to Reynolds numbers Re 1300 $<$ Re $<$ 63000 \cite{bodenschatz}, which is indeed in the turbulent regime. Upon analyzing the experimental results depicted in Fig.\,\ref{Fig-5}, the data set is properly fitted by making use of the hyperstatistics approach.
Note that the experimental results depicted in Fig.\,\ref{Fig-5} refer to an overlap of the data for various values of $R_{\lambda}$, being the obtained value of $q$ slightly modified upon analyzing distinct $R_{\lambda}$ values independently  (Supplementary Section 9).
Essentially, systems with inherent long-tailed decays embody finite probabilities of momenta with corresponding very low energy values, cf.\,Figs.\,\ref{Fig-3}, \ref{Fig-4}, and \ref{Fig-5}. As a consequence, the long-tailed decay process is influenced by the inclusion of low-probability energy events, which is naturally captured within our hyperstatistical framework (Supplementary Section 10).
In the following, we discuss the implications of hyperstatistics to the dielectric response function in non-homogeneous systems employing the here proposed $q$-$\gamma$-distribution for the relaxation time in the low-frequency response regime. 

\subsection{Dielectric response functions, power-law-type behaviour, and hyperstatistics}\label{dielectric}
In his seminal book, H. Frohlich already states that one must consider a distribution of relaxation times that contributes in the computation of both the real $\varepsilon_1$ and imaginary $\varepsilon_2$ part of the dielectric constant \cite{frohlich}. Such a distribution of relaxation times is considered by Frohlich based on  the fact that each individual electric dipole of the system has a particular associated relaxation time. However, the explicit mathematical form of the distribution function was not discussed by Frohlich. Before presenting our approach to the power-law-type behaviour of the dielectric response function using hyperstatistics, we recall the discussions concerning $\Gamma(n,q)$. It is clear that the Mellin transform of the $q$-exp function naturally leads to $\Gamma(n,q)$, cf.\,Eq.\,\ref{GammaMellin}, so that following Eq.\,S46 in the Supplementary Section 4, we can write:
\begin{equation}\label{nqGamma}
\int_0^\infty \tau^{n-1}\exp_q\left(-\frac{\tau}{\langle \tau \rangle}\right)\, d\tau = \langle \tau \rangle^{n}\Gamma(n,q).
\end{equation}
Although $\Gamma(n,q)$ is not explicitly employed in the fitting procedure of the experimental data set presented in this Section, it plays a structural role as the proper mathematical function to treat long-tailed decays, cf.\,Eq.\,\ref{nqGamma}, a key feature to treat systems of interest, e.g., the dielectric response functions in the regime $\tau \rightarrow \infty$, i.e., low-frequency response functions. In this regard, we recall Debye equations \cite{frohlich}:
\begin{equation}
\varepsilon_{1}(\omega) = \varepsilon_{\infty} +
\frac{\Delta \varepsilon}{1+\omega^{2}\tau^{2}},
\label{epsilon1}
\end{equation}
\begin{equation}
\varepsilon_{2}(\omega) = 
\frac{\Delta \varepsilon\,\omega\tau}{1+\omega^{2}\tau^{2}},
\label{epsilon2}
\end{equation}
with $\Delta \varepsilon = (\varepsilon_{s}-\varepsilon_{\infty})$ and $\omega$ the frequency. At this point, we make use of Eq.\,\ref{q-gamma2} written in terms of $\tau$, namely $f_q(\tau)$, given by:
\begin{equation}
f_q(\tau) = \frac{1}{\Gamma(n,q)}\left(\frac{n}{\langle \tau \rangle}\right)^n\tau^{n-1}\exp_q\left(-\frac{n\tau}{\langle \tau \rangle}\right).
\label{q-gamma}
\end{equation}
In the limit of $\tau \rightarrow \infty$, $f_q(\tau)$ reduces to:
\begin{equation}
f_q(\tau) \simeq \frac{1}{\Gamma(n,q)} \left(\frac{n}{\langle\tau\rangle}\right)^n \tau^{-1-\mu},
\label{fq}
\end{equation}
with $\mu = 1/(q-1) - n$. The admissible range of $\mu$ values imposed by the $q$-$\gamma$ distribution is in line with the causality constraints dictated by Kramers-Kronig relations.
Considering a distribution of $\tau$ given by Eq.\,\ref{fq}, and employing Eqs.\,\ref{epsilon1} and \ref{epsilon2}, we obtain \cite{gradsteyn}:
\begin{equation}
\varepsilon_1^* \simeq \int_0^\infty \frac{\Delta\varepsilon}{(1+\omega^2\tau^2)}\cdot\frac{1}{\Gamma(n,q)}\left(\frac{n}{\langle\tau\rangle}\right)^n\tau^{-1-\mu}d\tau,
\label{epsilon_1}
\end{equation}\\
\begin{equation}
\varepsilon_2 \simeq \int_0^\infty \frac{\Delta\varepsilon\omega\tau}{(1+\omega^2\tau^2)}\cdot\frac{1}{\Gamma(n,q)}\left(\frac{n}{\langle\tau\rangle}\right)^n\tau^{-1-\mu}d\tau,
\label{epsilon_2}
\end{equation}\\
where $\varepsilon_1^* = (\varepsilon_1 - \varepsilon_{\infty})$. After solving the integrals in Eqs.\,\ref{epsilon_1} and \ref{epsilon_2} and making some simplifications, we have:
\begin{equation}\label{epsilon3}
\varepsilon_1^{*} \simeq - \frac{\Delta \varepsilon}{\Gamma(n,q)} \cdot \left(\frac{n}{\langle\tau\rangle} \right)^n \cdot \frac{\pi}{2} \csc(\mu \pi / 2) \cdot \omega^\mu,
\end{equation}

\begin{equation}\label{epsilon4}
\varepsilon_2 \simeq \frac{\Delta \varepsilon}{\Gamma(n,q)} \cdot \left(\frac{n}{\langle\tau\rangle} \right)^n \cdot \frac{\pi}{2} \sec (\pi \mu/2) \cdot \omega^\mu,
\end{equation}
with $0 < \mu < 1$. Using literature results for $\mu$ \cite{jonscher1977} and considering $n = 1$, we deduce $q$-values for the dielectric response functions ranging from $\sim$1.71 to $\sim$1.96, which are consistent with the range of possible $q$-values imposed by $\Gamma(n,q)$, cf.\,Eq.\,\ref{inequality}, considering its limit of convergence.  Note that $\Gamma(n,q)$ governs the domain of physically acceptable range of $n$ and $q$, normalization, and asymptotic behaviour of the relaxation-time distribution. We demonstrate that the emergence of power-law dielectric response follows from the long-time ($\tau \rightarrow \infty$) tail of the $f_q(\tau)$ relaxation-time PDF. Thus, we can infer that the observed scaling laws arise as a direct consequence of the underlying $f_q(\tau)$ PDF of $\tau$, giving rise to criticality classes in terms of $q$. Note that a similar analysis can also be carried out for the dielectric constant and electrical conductivity for the Drude model \cite{ashcroft}.

In practice, to make use of our approach, the data set should present a long-tail decay not captured by standard exponential and $q$-exp functions. The data set of interest should be fitted by employing either the $(q, n)$-generalized Boltzmann factor $B_q^n$, cf.\,Eq.\,\ref{qBfactor}, for mesoscopic (local, i.e., at the domain level) measurements, or $B_H$, cf.\,Eq.\,\ref{B_H(E)}, for macroscopic (global) ones. The computation of $\langle\tau\rangle$ shall be different depending on the particular PDF governing the system. Upon fitting the data, one should check whether the obtained values of $q$ and $n$ lie within the convergence regime of the $\Gamma(n,q)$ function, cf.\,Fig.\,\ref{Fig-1}. Hence, by making use of this simple protocol, it is possible to extract the value of $q$ and $n$ of the particular data set of interest and thus to infer \emph{universality classes} in terms of such parameters. Yet, note that for $q < 1$ the asymptotic long-tailed behaviour of the $q$-exp function is suppressed, cf.\,top-right panel of Fig.\,3.1 of Ref.\,\cite{Tsallisbook}, which places this regime outside of the validity scope of hyperstatistics. Also, since hyperstatistics is based on the generalization of the $\Gamma(n,q)$ function, its convergence criterion must be obeyed, cf.\,Eq.\,\ref{inequality}. Hence, values of $q$ $<$ 1 do not ensure a convergent integral of $\Gamma(n,q)$, cf.\,Eqs.\,\ref{Gamma nq} and \ref{inequality}.

\section{Distinguishing hyperstatistics from other approaches}\label{distinguishing}
The distinction between superstatistics and hyperstatistics should be understood primarily in terms of the hierarchical level at which the statistical description is introduced and in terms of the mathematical existence of a particular functional form, cf.\,discussions below. Indeed, the mathematical construction of an effective statistical weight, $B(E)=\int_0^\infty f(\beta)\exp(-\beta E)\,d\beta$, is a general construction and is not a mathematical property originating exclusively from Beck and Cohen. Regarding the consideration of a distribution of $\beta$, the fingerprint of superstatistics, it is worth recalling that G. Wilk and Z. W{\l}odarczyk in Ref.\,\cite{wilk2000} have already reported that $q > 1$ is given ``\emph{... entirely by the fluctuations of the parameters of the usual exponential distribution. ...}''. \newline Based on the fact that the $\gamma$-distribution incorporates a distribution of exponential weights (Supplementary Section 3), as well as based on experimental observations \cite{eli1,eli2,eli3,afewremarks}, we have assumed that we are dealing with slightly distinct Boltzmann factors inside a single domain, cf.\,Fig.\,\ref{Fig-2}. Thus, we compute the expected value of the Boltzmann factor inside a single domain, which leads to a \emph{local} non-Boltzmann-Gibbsian statistics described by a $q$-exp-based $(q, n)$-generalized Boltzmann factor, cf.\,Eq.\,\ref{qBfactor}. Indeed, the expected value of a mathematical function $F(x)$, cf.\,Eq.\,(3.2) of Ref.\,\cite{mahon}, is given by:
\begin{equation}
\langle F(x) \rangle = \int_{-\infty}^{\infty}dx p(x)F(x),
\label{mahon}
\end{equation}
where $F(x)$ is itself a random variable and $p(x)$ is the PDF considered, see also Eq.\,(12.2) of Ref.\,\cite{frohlich}. Of course, the lower limit of the integral in Eq.\,\ref{mahon} should be adapted to the situation in hand. At this point one should be very careful because the integration limits are crucial, since finite integration limits preclude the consideration of a genuine Laplace transform, cf.\,Supplementary Section\,7.1. As a matter of fact, in hyperstatistics $F(x) = e^{-\beta_i E}$ and the integration interval [0, $\infty$], so that for the $\gamma$-distribution a $q$-exp is achieved as $B_q^n(E) = \langle \exp(-\beta_i E) \rangle = \textmd{exp}_q(-\langle \beta_i \rangle E)$, cf.\,Eq.\,\ref{qBfactor}. In superstatistics, the $q$-exp is considered to emerge from a distribution of inverse temperatures in the whole system \cite{cbeckfig}. In hyperstatistics, by contrast, the $q$-exp is interpreted as an effective statistical weight arising from a distribution of Boltzmann factors at the domain level, i.e., mesoscopic scale, cf.\,Fig.\,\ref{Fig-2}, where BG statistics is no longer assumed to be valid. The fact that $B_q^n(E)$ incorporates a $q$-exp function, indicates that we have moved locally from Boltzmann-Gibbsian to $q$-statistics space. This is in line with experimental observations, e.g., as in the case of electrodes attached to a person’s head to record electroencephalogram data \cite{afewremarks}, whole body perturbations at cell level obtained via a deep learning method \cite{erturk2026}, or in temperature measurements \cite{eli1,eli2,eli3}.  Another crucial difference between the two approaches refers to the estimation of the expected value of the observable. While superstatistics assumes  the expected value of $\beta$ as $\langle \beta \rangle = \int_0^{\infty}\beta f(\beta)d\beta$ \cite{beck2003}, which is a well-known procedure from textbooks \cite{mahon,frohlich}, hyperstatistics in turn considers not only the $q$-generalized $\gamma$-distribution introduced in this work, but also the fact that time $t$ and $RC$ are conjugate variables, analogous to pressure and volume in thermodynamics \cite{lorentz}. As a matter of fact, in a real capacitor, it is natural to expect local, i.e., at the domain level, variations of both $R$ and $C$ in the dielectric medium, so we propose that it is meaningful to estimate $\langle \tau \rangle$ as follows (Supplementary Section 5):
\begin{equation}
\langle \tau \rangle = \int_0^{\infty}\tau q\textmd{-}\gamma(1/\tau)d(1/\tau) = \frac{(n\tau_0)[1+n(1-q)]}{(n-1)}.
\label{1/tau}
\end{equation}
\begin{table}[t]
\centering
\renewcommand{\arraystretch}{1.7}
\setlength{\tabcolsep}{6pt}
\renewcommand{\tabularxcolumn}[1]{m{#1}}
\begin{tabularx}{\linewidth}{|>{\hsize=1.12\hsize\centering\arraybackslash}X|>{\hsize=0.88\hsize\centering\arraybackslash}X|}
\hline
\textbf{Hyperstatistics} & \textbf{Superstatistics for the $\gamma$ PDF} \\
\hline
Non-Boltzmann--Gibbsian domains
  & Boltzmann--Gibbsian domains \\
\hline
($q$, $n$)-generalized Boltzmann factor
  & $q$-generalized Boltzmann factor \\
\hline
Two independent fitting parameters $q$ and $n$: $1 < q < (1 + 1/n)$
  & Single fitting parameter $q$, which is tied to $n$ by $q = (1 + 1/n)$ \\
\hline
\rule[-4ex]{0pt}{9ex}\mbox{$\displaystyle \langle \beta_i \rangle = \int_0^{\infty}\beta_i\, q\textmd{-}\gamma(\beta_i)\, d(\beta_i) = \frac{\beta_0}{[1+(1-q)(n+1)]}$}
  & \mbox{$\displaystyle \langle \beta \rangle = \int_0^{\infty}\beta\, f(\beta)\, d\beta = \beta_0$} \\
\hline
$(q,n)$-dependent $\langle \beta_i \rangle$ incorporated in $B_q^n(E)$
  & $\langle \beta \rangle = \beta_0$ enters $B(E)$ only as a fixed value, independent of $q$ \\
\hline
$\Gamma(n,q)$ dictates the convergence criterion of $B_q^n(E)$ and the range of validity of $\langle \beta_i \rangle$
  & $-$ \\
\hline
$\,B_H(E) = \frac{{}_2F_1(m, n; 2m; 1 - b^*E)}{{}_2F_1(m, n; 2m; 1)}\,$, with $m = 1/(q-1)$;
\newline\newline
$\lim_{m\rightarrow\infty} B_H(E) = \exp_q(-\langle \beta_0 \rangle E)$;
\newline\newline
\vspace{0.15cm}$\lim_{n\rightarrow\infty}\lim_{m\to\infty} B_H(E) = \exp(-\langle\beta_0 \rangle E)$
  & $B(E) = \exp_q(-\beta_0 E)$ \\
\hline
\end{tabularx}
\caption{Comparison between hyperstatistics and superstatistics for the $\gamma$ PDF.}
\label{hyperandsuper}
\end{table}
This leads to $\langle \tau \rangle$ that depends explicitly on both $q$ and $n$. Note that superstatistics is pinned in the regime where $q = (1 + 1/n)$, cf.\,solid black line in Fig.\,\ref{Fig-1} b), while hyperstatistics enables us to explore the whole range $1 < q < (1 + 1/n)$, which obeys the convergence criterion of $\Gamma(n, q)$, cf.\,Eqs.\,\ref{Gamma nq} and \ref{inequality}. As a result, hyperstatistics incorporates $q$ and $n$ as independent parameters, which in turn provides a higher accuracy when fitting a data set of interest without the need to change the PDF considered to obtain $B_q^n$. The improved agreement with experimental results should therefore be regarded as a strong experimental indication supporting the hyperstatistics hierarchy, i.e., \emph{local} non-Boltzmann-Gibbsian statistics, rather than simply as a consequence of using a more flexible fitting function. The values of $q$ and $n$ obtained in the fittings of the various experimental data sets give us physical insights on the deviation of a canonical exponential decay and the inherent degree of heterogeneity, respectively. Remarkably, for $q \rightarrow 1$ and $n \rightarrow \infty$ $\Rightarrow \langle \tau \rangle \rightarrow \tau_0$. The distinctions between hyperstatistics and superstatistics therefore concern not merely the resulting $q$-exp weight, but the underlying distribution and its domain-level physical interpretation, cf.\,Table\,\ref{hyperandsuper}.
Hence, despite sharing some similarities, hyperstatistics and superstatistics are fundamentally distinct frameworks, differing in their underlying assumptions, and physical interpretation. It is clear that neither a simple exponential nor a $q$-exp function is sufficient to capture such intricate local non-Boltzmann-Gibbsian behavior, as these functions fail to capture the key physical mechanisms underlying such a behavior, unlike $B_q^n(E)$, as can be seen in the fits depicted in Fig.\,\ref{Fig-5}.
\section{Conclusions and outlook}\label{Conclusions}
We have introduced the $\Gamma(n,q)$ function as the Mellin transform of the $q$-exp function. We have fixed the recurrence relation and convergence conditions of $\Gamma(n,q)$. As the highlight of the present work, we have proposed a new framework, named by us hyperstatistics, upon considering non-Boltzmann-Gibbsian-type statistics at the domain level of the system, whose corresponding entropic functional is the one proposed in Ref.\,\cite{tsallis1988}. We have shown that hyperstatistics and superstatistics are equivalent at the domain level. At the macroscopic level, hyperstatistics recovers superstatistics for $q \to 1$, since the domains become Boltzmann-Gibbsian, being also BG statistics recovered for $n \to \infty$.
We have probed our approach  experimentally upon considering a $\gamma$-type distribution of relaxation times in the discharge process of a real capacitor in a standard series RC circuit, the pumping of the $^4$He circulation lines in a cryostat, midrapidity data set for \emph{p}-Pb collisions at the LHC, as well as acceleration in turbulent systems. Also, we have deduced the power-law-type dielectric response function of non-homogeneous systems \cite{jonscher1977, bohmer1993}. 
Beyond the specific examples discussed here, hyperstatistics provides a unified description for relaxation and response functions in complex systems where BG statistics is no longer applicable. This includes critical phenomena, response functions, e.g., dielectric response, non-homogeneous systems, among others. Our approach can be also employed to other scenarios, such as in cases involving non-Gaussian distributions, e.g., temperature fluctuations in the cosmic microwave background \cite{bernui}, luminosity calibration in charged particle colliders \cite{abed}, power-law in both velocity distributions \cite{benetti} and trapped ions \cite{devoe}, as well as for the standard map \cite{ugursr} and quantum decays \cite{lagnese2024, coleman1977, kofman2000}. The obtaining of the entropic functional for $B_H$ constitutes part of future works.

\section*{Methods}\label{sec11}
The RC electrical circuit consists of a resistor with $R$ = 1\,M$\Omega$ and an electrolytic aluminium capacitor of $C$ = 470\,$\mu$F supplied by JWCO group, so that the nominal relaxation time is 470\,s. We have employed a Keithley 617 electrometer as a voltage source and a nanovoltmeter Keithley 2182A. 
The data set shown in the bottom panel of Fig.\,\ref{Fig-3} was recorded using a MercuryiTC controller provided by Oxford Instruments and the pumping was carried out employing a Hi-Cube vacuum pump supplied by Pfeiffer attached to the $^4$He circulation lines of a Teslatron-PT cryostat from Oxford Instruments. 

\bmhead{Acknowledgements}
MdeS acknowledges partial financial support from the S\~ao Paulo Research Foundation -- Fapesp (Grants 2011/22050-4, 2017/07845-7, and 2019/24696-0), National Council of Technological and Scientific Development -- CNPq (Grants 303772/2023-9), and Capes -- Finance Code 001 (M.\,Sc.\,fellowship of SMS). CT acknowledges partial financial support from National Council of
Technological and Scientific Development -- CNPq and Funda\c{c}\~ao Carlos Chagas Filho de Amparo \`a Pesquisa do Estado do Rio de Janeiro -- Faperj. MdeS acknowledges Christian Beck for fruitful suggestions, including the mentioning of the paper by G.E. Crooks regarding the terms ``ensemble of ensembles'' and hyperensemble. We thank the anonymous Reviewers for their thoughtful and constructive comments, which helped us to improve the quality and clarity of the manuscript.

\bmhead{Data availability}
The data that support the findings of this work are openly available \cite{zenodo}.

\bmhead{Author contributions}
S.M.S., L.S., and M.deS. carried out the RC circuit and $^4$He lines pumping experiments. 
M.deS. carried out the calculations with inputs from S.M.S. and L.S.. S.M.S. and L.S. carried out the data fits. M.deS. wrote the paper with contributions from S.M.S. and L.S.. C.T. created $q$-statistics, which served as the ground for hyperstatistics. All authors revised the manuscript. M.deS. conceived and supervised the whole project.

\bmhead{Competing interests}
The authors declare no competing interests.

\end{document}